\documentclass[aps,physrev,twocolumn,superscriptaddress]{revtex4}
\usepackage{graphicx}
\usepackage{siunitx}
\usepackage{bm}
\usepackage[caption=false]{subfig}
\usepackage{mathtools}
\usepackage{multirow}
\usepackage{amssymb}
\usepackage{braket}
\usepackage{amsmath}
\usepackage{placeins}
\usepackage{cancel}
\usepackage{xcolor}
\usepackage{spreadtab}
\usepackage{numprint}

\begin{document}

\title{Structural, electronic, elastic, power and transport properties of $\beta$-Ga$_2$O$_3$ from first-principles}

\author{Samuel Ponc\'e}
\email{samuel.ponce@epfl.ch}
\affiliation{%
Theory and Simulation of Materials (THEOS), \'Ecole Polytechnique F\'ed\'erale de Lausanne,
CH-1015 Lausanne, Switzerland
}%
\affiliation{%
Department of Materials, University of Oxford, Parks Road, Oxford, OX1 3PH, UK
}%
\author{Feliciano Giustino}
\email{fgiustino@oden.utexas.edu}
\affiliation{%
Oden Institute for Computational Engineering and Sciences, The University of Texas at Austin, Austin, Texas 78712, USA
}%
\affiliation{%
Department of Physics, The University of Texas at Austin, Austin, Texas 78712, USA
}%
\affiliation{%
Department of Materials, University of Oxford, Parks Road, Oxford, OX1 3PH, UK
}%

\date{\today}

\begin{abstract}
We investigate the structural, electronic, vibrational, power and transport properties of the $\beta$ allotrope of Ga$_2$O$_3$ from first-principles.  
We find phonon frequencies and elastic constants that reproduce the correct band ordering, in agreement with experiment. 
We use the Boltzmann transport equation to compute the intrinsic electron and hole drift mobility and obtain a room temperature values of 258~cm$^2$/Vs and 1.2~cm$^2$/Vs, respectively as well as 6300~cm$^2$/Vs and 13~cm$^2$/Vs at 100~K. 
Through a spectral decomposition of the scattering contribution to the inverse mobility, we find that multiple longitudinal optical modes of B$_u$ symmetry are responsible for the electron mobility of $\beta$-Ga$_2$O$_3$ but that many acoustic modes also contributes, making it essential to include all scattering processes in the calculations.
Using the von Hippel low energy criterion we computed the breakdown field to be 5.8~MV/cm at room temperature yielding 
a Baliga's figure of merit of 1250 with respect to silicon, ideal for high-power electronics.
This work presents a general framework to predictively investigate novel high-power electronic materials.    
\end{abstract}

\maketitle

\section{Introduction}

The $\beta$ allotrope of Ga$_2$O$_3$ has attracted some attention as an ultra wide bandgap transparent semiconducting oxide~\cite{Galazka2018}.
As a consequence of its large bandgap, $\beta$-Ga$_2$O$_3$ possess a very high breakdown electric field of 8~MV/cm~\cite{Higashiwaki2012} and a large Baliga's figure of merit (BFOM)~\cite{Baliga1982}, which makes it a promising alternative to GaN and SiC for high-power electronics~\cite{Higashiwaki2018,Kong2019}. 
In addition it can be synthesized by melt-growth method which allows for low cost and large-scale production~\cite{Villora2004,Aida2008}.
Its electronic and optical properties also make it a good candidate for UV transparent conducting oxide (TCO)~\cite{Nomura2004,Furthmueller2016}.

One property of $\beta$-Ga$_2$O$_3$ that makes it so attractive is its high carrier mobility for a material with such a wide bandgap. 
The electron mobility of $\beta$-Ga$_2$O$_3$ has been studied more extensively than the hole mobility due to experimental interest and the fact the hole mobility is two order of magnitude smaller. 
Given the promise offered by $\beta$-Ga$_2$O$_3$, it is surprising that many basic properties have not been investigated in detail. 
From a theoretical perspective, this might be due to the fact that $\beta$-Ga$_2$O$_3$ has a 10-atom primitive cell, which makes first-principles calculations in this material more challenging than for standard tetrahedral semiconductors. 
In particular, the shape of the conduction band was not well understood until recently.
Indeed, Ueda~\textit{et al.}~\cite{Ueda1997} measured a strong anisotropy of the conduction-band effective mass.
However, since then many experiments and theoretical studies indicated that the conduction band is nearly isotropic~\cite{Yamaguchi2004,Irmscher2011,Villora2004,Varley2010,Varley2011,Peelaers2015,Wong2016,Ghosh2020}.
Another question relates to the relative importance of nonpolar optical-phonon, polar optical-phonon, and ionized-impurity scattering at room temperature.
Initially it was thought that the dominant scattering mechanism in $\beta$-Ga$_2$O$_3$ was due to nonpolar optical phonons with a large deformation potential of 4$\times$10$^9$~eV/cm~\cite{Parisini2016}.
However, Ghosh and Singisetti~\cite{Ghosh2016} identified a longitudinal-optical phonon mode with energy around 21~meV as the dominant mechanism in the mobility of $\beta$-Ga$_2$O$_3$, and this finding was later confirmed by multiples authors~\cite{Ma2016,Kang2017,Mengle2019}.
Finally, there is some debate about the ordering of the zone-centred phonons, namely the Raman-active A$_g$ mode and the infrared-active B$_u$ TO mode~\cite{Machon2006,Liu2007,Schubert2016,Mengle2019}.

One crucial material property for high-power electronics is the breakdown field, i.e. the magnitude of the external electric field that a material can sustain before incurring permanent damage. 
The breakdown field can be computed from first-principles using the von Hippel low energy criterion
~\cite{VonHippel1937,Sparks1981,Sun2012a}, and was recently computed \textit{ab-initio} by Mengle and Kioupakis~\cite{Mengle2019} to be 5.4~MV/cm in $\beta$-Ga$_2$O$_3$ considering only the dominant LO phonon mode. 
They further estimated that considering all modes would increasing the theoretical intrinsic breakdown field by 20\% to 6.8~MV/cm.
Such calculation assumes total impact ionization for all electrons with energies above the bandgap, and should therefore be
seen as a lower bound; it can also be improved by computing impact ionization coefficient from first principles~\cite{Ghosh2018}.

The BFOM~\cite{Baliga1982} describes the current handling capability of a material and is often given relative to silicon. 
In addition to the breakdown field, the second material's parameter entering into the BFOM is the intrinsic carrier mobility. 
The electron room-temperature mobility of $\beta$-Ga$_2$O$_3$ was computed to be 115~cm$^2$/Vs at a carrier concentration of $10^{17}$~cm$^{-3}$, with a temperature dependence in good agreement with experiment~\cite{Ghosh2016}, by using Wannier interpolation of the electron-phonon matrix elements~\cite{Ponce2016a} and Rode's method~\cite{Rode1975}.  
The mobility was also estimated to be below 200~cm$^2$/Vs using $\mathbf{k}\cdot\mathbf{p}$ perturbation theory~\cite{Ma2016}.

In this context, a careful and detailed analysis of the crystal structure, electronic, optical, vibrational, elastic, and transport properties of $\beta$-Ga$_2$O$_3$ is warranted. 
The manuscript is organized as follows. 
In Section~\ref{sec:structure} we discuss the relaxed crystal structure of the monoclinic $\beta$-Ga$_2$O$_3$ and the importance of spin-orbit coupling.
Section~\ref{sec:electronic} is dedicated to the study of the electronic properties including bandgaps, electronic bandstructure and effective masses.
In Section~\ref{sec:vibration} we analyze the phonon dispersion, infrared and Raman spectra, dielectric constant and Born charges, and elastic properties.  
The Section~\ref{sec:mob} present the computed electron and hole carrier mobility with temperature as well as a mode-resolved analysis of the scattering contribution to the mobility.
Finally in Section~\ref{sec:fom}, we discuss and compute Baliga's figure of merit of $\beta$-Ga$_2$O$_3$ and compare it with Silicon.

\section{Crystal Structure}\label{sec:structure}
The crystal structure of $\beta$-Ga$_2$O$_3$ was originally investigated by Geller to be monoclinic with the C$_{2h}$ (2/m) point group~\cite{Geller1960} and later refined by {\AA}hman~\cite{Ahman1996} using single crystal diffraction.
The measured lattice parameters of the conventional unit cell are $a$=12.214~\AA, $b$=3.037~\AA, $c$=5.798~\AA\,and $\beta$=103.83$^\circ$~\cite{Ahman1996}. 
The conventional cell vectors are $(a,0,0)$, $(0,b,0)$ and $(c\cos\beta,0,c\sin\beta)$ while the primitive cell vectors are  $(\frac{a}{2},-\frac{b}{2},0)$, $(\frac{a}{2},\frac{b}{2},0)$ and $(c\cos\beta,0,c\sin\beta)$. 
Any atomic coordinate expressed in the conventional cell $(c_x,c_y,c_z)$ can be expressed in the primitive cell by using the transformation $(c_x-c_y,c_x+c_y,c_z)$.
The primitive and conventional cell are made of 10 and 20 atoms, respectively.

The gallium atom sits in two inequivalent position with octahedral and tetrahedral coordination, respectively. 
There are three inequivalent oxygen atoms occupying a distorted cubic lattice, with two oxygen atoms being three-fold coordinated and one oxygen atom fourfold coordinated. 
All the five inequivalent atoms have $4i$ Wyckoff position which corresponds to symmetry ($x$,$0$,$z$) and ($-x$,$0$,$-z$). 
The system has four crystal symmetries: the identity, a $\pi$ rotation around the Cartesian $y$ axis
and their inversions.

To determine the atom positions, we relaxed the lattice parameters and atomic coordinates, starting from the experimental data.
We used the Quantum ESPRESSO software suite~\cite{Giannozzi2017} with relativistic LDA pseudopotentials from PseudoDojo~\cite{Setten2018}, including the 3$s^2$ 3$p^6$ 3$d^{10}$ 4$s^2$ 4$p^1$ semicore states for gallium and the 2$s^2$ 2$p^4$ electrons for oxygen. 
The wavefunctions were expanded in a plane-wave basis set with energy cutoff of 120~Ry (160~Ry for the elastic response) and an homogeneous $\Gamma$-centered Brillouin-zone sampling of $8\times 8\times 8$ points.
We converged the structure such that the maximum force was smaller than 2$\cdot$10$^{-7}$~Ry/\AA\,and the maximum stress component was lower than 0.07~Ry/\AA$^3$.

\begin{figure}[t]
  \centering
  \includegraphics[width=0.99\linewidth]{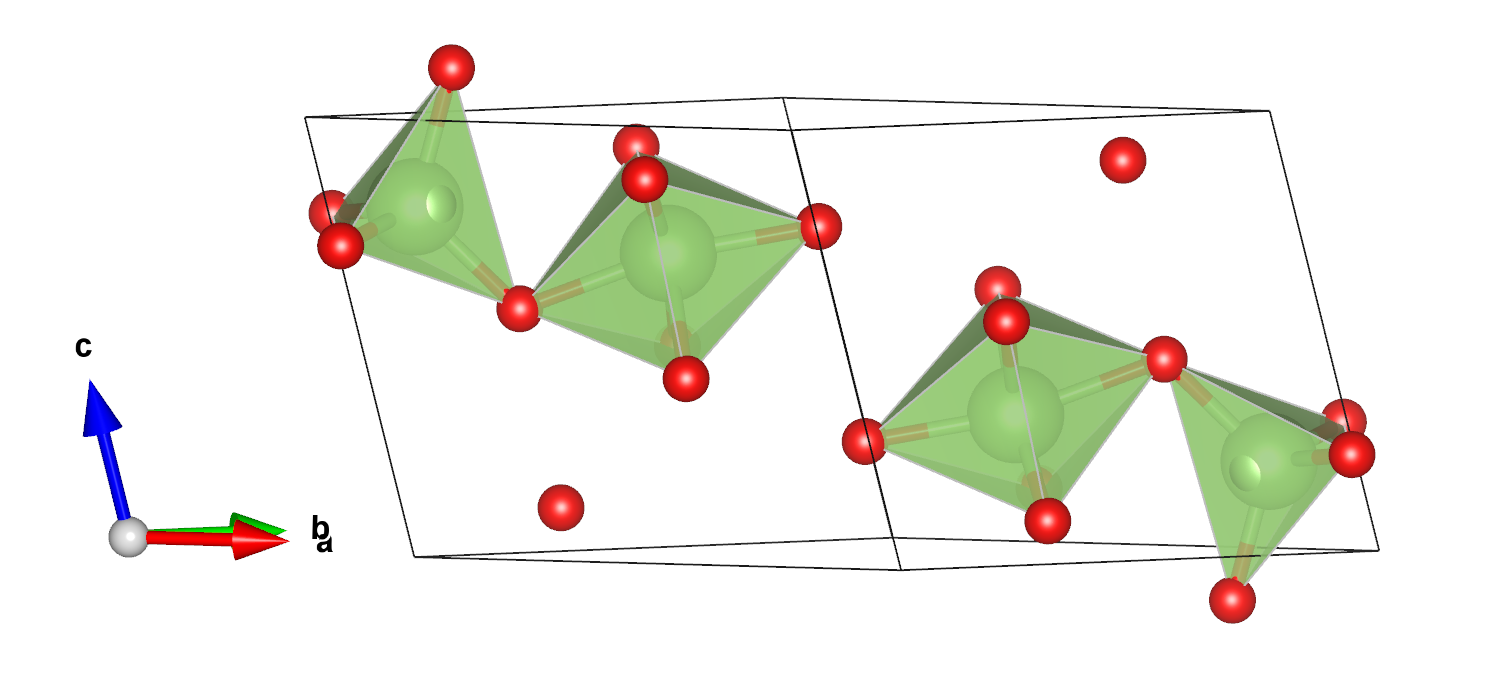}
  \caption{\label{fig1}
   Relaxed crystal structure of the primtive cell of $\beta$-Ga$_2$O$_3$ where the large atoms are gallium and the small red atoms are oxygen. Rendered using VESTA~\cite{Momma2011}.
  }
\end{figure}
The relaxation yielded the lattice parameters $a$=12.128~\AA, $b$=3.016~\AA, $c$=5.752~\AA\,and $\beta$=103.75$^\circ$, which slightly underestimates the experimental one as expected from LDA. 
The relaxed primitive cell crystal structure is shown in Fig.~\ref{fig1} and is formed by two distorted octahedra and two distorted tetrahedra. 
The gallium and oxygen atoms occupy two and three inequivalent sites at the $4i$ Wyckoff position, respectively, whose coordinates are provided in Table~\ref{table1} and are in close agreement with the experimental assignment~\cite{Ahman1996}. 
The inequivalent gallium-oxygen bond lengths are also reported in Table~\ref{table1} with the thetrahedra having smaller bond-lengths than the octahedra. Interestingly, due to their distorted nature, there are two inequivalent Ga$_{\rm II}$-O$_{\rm III}$ bond lengths in the octahedral configuration despite having only one inequivalent oxygen position.

We also report in Table~\ref{table1} the volume, density, atomic coordinates, and bond lengths, and compare them with experimental data. 
The calculations were made without spin-orbit coupling (SOC) but we tested that including this effect modifies the crystal data shown in Table~\ref{table1} by less than 0.005\%.
Hence, this effect is neglected for the rest of this work. 
We finally note that the primitive cell vectors can equivalently be rotated such that $a$=$b$=11.809~\AA, $c$=10.869~\AA, $\alpha=\beta=103.335^{\circ}$ and $\gamma=27.933^{\circ}$.

\npdecimalsign{.}
\nprounddigits{6}
\begin{table}[ht]
  \begin{tabular}{l n{3}{6} n{3}{6} | n{3}{6} n{3}{6} }
  \toprule\\
Lattice       &  \multicolumn{2}{r}{Experimental~\cite{Ahman1996}} &  \multicolumn{2}{r}{This work}   \\   
a (\AA)       & \multicolumn{2}{r}{12.214}        & \multicolumn{2}{r}{12.128}\\
b (\AA)       & \multicolumn{2}{r}{3.037}         & \multicolumn{2}{r}{3.016}\\
c (\AA)       & \multicolumn{2}{r}{5.798}        & \multicolumn{2}{r}{5.752}\\
$\beta$ ($^{\circ}$) & \multicolumn{2}{r}{103.83} & \multicolumn{2}{r}{103.750}  \\
\multicolumn{2}{l}{Volume (\AA$^3$)}  &   \multicolumn{1}{r}{104.425} & \multicolumn{2}{r}{102.190} \\
\multicolumn{2}{l}{Density (g/cm$^3$)} & \multicolumn{1}{r}{5.961} & \multicolumn{2}{r}{6.09171} \\[0.5em]
       &  \multicolumn{4}{c}{Coordinates}   \\  
       &  \multicolumn{2}{c}{Experimental} &  \multicolumn{2}{c}{This work}   \\  
Atom   &   \multicolumn{1}{c}{$x$}  &   \multicolumn{1}{c}{$z$}  & \multicolumn{1}{c}{$x$}  &   \multicolumn{1}{c}{z} \\
\hline 
Ga$_{\rm I}$ ($4i$)  & 0.09050 &  0.79460 & 0.0907608352 & 0.7951120184 \\
Ga$_{\rm II}$ ($4i$) & 0.34134 &  0.68598 & 0.3411450768 & 0.6855889459 \\
O$_{\rm I}$  ($4i$)  & 0.16450 &  0.10980 & 0.1655995553 & 0.1094726070 \\
O$_{\rm II}$ ($4i$)  & 0.49590 &  0.25660 & 0.4963406016 & 0.2557307370 \\
O$_{\rm III}$ ($4i$) & 0.82670 &  0.43680 & 0.8269104067 & 0.4378690351 \\
           \multicolumn{5}{c}{ }   \\[-0.5em]
         &  \multicolumn{4}{c}{Distances (\AA) }   \\
  Pairs  &  \multicolumn{2}{r}{Experimental~\cite{Ahman1996}} &  \multicolumn{2}{r}{This work}   \\   
\hline  
Ga$_{\rm I}$-O$_{\rm I}$    &  \multicolumn{2}{r|}{1.835} & \multicolumn{2}{r}{1.818}  \\ 
Ga$_{\rm I}$-O$_{\rm II}$   &  \multicolumn{2}{r|}{1.833} & \multicolumn{2}{r}{1.825}  \\ 
Ga$_{\rm I}$-O$_{\rm III}$  &  \multicolumn{2}{r|}{1.863} & \multicolumn{2}{r}{1.852}  \\ 
Ga$_{\rm II}$-O$_{\rm I}$   &  \multicolumn{2}{r|}{1.937} & \multicolumn{2}{r}{1.924}  \\ 
Ga$_{\rm II}$-O$_{\rm II}$  &  \multicolumn{2}{r|}{1.935} & \multicolumn{2}{r}{1.919}  \\ 
Ga$_{\rm II}$-O$_{\rm III}$ &  \multicolumn{2}{r|}{2.005} & \multicolumn{2}{r}{1.992}  \\
Ga$_{\rm II}$-O$_{\rm III}^{\dagger}$ &  \multicolumn{2}{r|}{2.074} & \multicolumn{2}{r}{2.054}  \\ 
  \botrule     
  \end{tabular}
  \caption{\label{table1}
 Experimental and LDA crystal atom coordinates of $\beta$-Ga$_2$O$_3$ without spin-orbit coupling with conventional unit cell coordinates $(0 0 0, \frac{1}{2}  \frac{1}{2} \frac{1}{2})\pm(x 0 z)$ (20 atoms) and primitive cell 
 coordinates $\pm(x x z)$ (10 atoms). In parenthesis after the non-equivalent atoms, we indicate their Wyckoff position. 
 $^{\dagger}$ There are two inequivalent bond length Ga$_{\rm II}$-O$_{\rm III}$ in the octahedral configuration. 
  }
\end{table}

\section{Electronic properties}\label{sec:electronic}

\npdecimalsign{.}
\nprounddigits{2}
\begin{table}
  \begin{tabular}{l n{5}{2} n{1}{2} n{1}{2} n{1}{2} n{1}{2} n{1}{2}}
  \toprule\\
        &  \multicolumn{1}{c}{Direct gap (eV)} &  \multicolumn{5}{c}{Indirect gap(eV)}  \\ 
This work LDA                    & 2.5532 & \multicolumn{5}{c}{2.53} \\ 
LDA~\cite{Yamaguchi2004}         & 2.188  & \multicolumn{5}{c}{-} \\ 
GGA-AM05~\cite{Furthmueller2016} & 2.377  & \multicolumn{5}{c}{2.36} \\  
HSE+G$_0$W$_0$~\cite{Furthmueller2016} & 5.038  & \multicolumn{5}{c}{5.05} \\       
HSE06~\cite{Peelaers2015}        & 4.88   & \multicolumn{5}{c}{4.84} \\
Experiment~\cite{Orita2000}      & \multicolumn{1}{c}{-}      & \multicolumn{5}{c}{4.90} \\ 
Experiment~\cite{Matsumoto1974}  & \multicolumn{1}{c}{-}      & \multicolumn{5}{c}{4.54} \\ 
Experiment~\cite{Tippins1965}    & \multicolumn{1}{c}{-}      & \multicolumn{5}{c}{4.70} \\   
\hline 
        &  \multicolumn{1}{c}{electron ($m_e$)} &  \multicolumn{5}{c}{Hole ($m_e$)}  \\  
        &        & \multicolumn{1}{c}{$\Gamma$X} & \multicolumn{1}{c}{$\Gamma$Y} & \multicolumn{1}{c}{$\Gamma$Z} & \multicolumn{1}{c}{IL$_\parallel$} & \multicolumn{1}{c}{IL$_\perp$} \\
This work LDA                 & 0.255 & \multicolumn{1}{c}{[-78]} & \multicolumn{1}{c}{3.40} & \multicolumn{1}{c}{0.35} & \multicolumn{1}{c}{3.0} & \multicolumn{1}{c}{3.6} \\
LDA~\cite{Yamaguchi2004}      & 0.24  & \multicolumn{1}{c}{-} & \multicolumn{1}{c}{-} & \multicolumn{1}{c}{-} & \multicolumn{1}{c}{2.90} & \multicolumn{1}{c}{4.19}  \\
HSE06~\cite{Mohamed2010}      & 0.28  & \multicolumn{1}{c}{-} & \multicolumn{1}{c}{-} & \multicolumn{1}{c}{-} & \multicolumn{1}{c}{-} & \multicolumn{1}{c}{-} \\
HSE06~\cite{Peelaers2015}     & 0.275 & \multicolumn{1}{c}{-} & \multicolumn{1}{c}{-} & \multicolumn{1}{c}{-} & \multicolumn{1}{c}{-} & \multicolumn{1}{c}{-} \\
B3LYP~\cite{He2006}           & 0.342 & \multicolumn{1}{c}{-} & \multicolumn{1}{c}{-} & \multicolumn{1}{c}{-} & \multicolumn{1}{c}{-} & \multicolumn{1}{c}{-} \\
HSE~\cite{Furthmueller2016}   & 0.268 & \multicolumn{1}{c}{-} & \multicolumn{1}{c}{-} & \multicolumn{1}{c}{-} & \multicolumn{1}{c}{-} & \multicolumn{1}{c}{-} \\
Experiment~\cite{Mohamed2010} & 0.28  & \multicolumn{1}{c}{-} & \multicolumn{1}{c}{-} & \multicolumn{1}{c}{-} & \multicolumn{1}{c}{-} & \multicolumn{1}{c}{-} \\
Experiment~\cite{Janowitz2011}& 0.28  & \multicolumn{1}{c}{-} & \multicolumn{1}{c}{-} & \multicolumn{1}{c}{-} & \multicolumn{1}{c}{-} & \multicolumn{1}{c}{-} \\
  \botrule     
  \end{tabular}
  \caption{\label{table2}
 Effective masses of $\beta$-Ga$_2$O$_3$. The electron effective mass is almost isotropic so we report the average, as done in prior studies. The hole effective mass is reported at $\Gamma$ as well as at the valence band maximum located between the I and L high-symmetry point. 
 The negative value between bracket is not an effective mass since the band curvature is positive; this value is reported as a measure of the band curvature.
  }
\end{table}

The room-temperature optical bandgap of $\beta$-Ga$_2$O$_3$ obtained through absorption measurements is estimated to be between 4.54~eV and 4.90~eV~\cite{Matsumoto1974,Orita2000,Tippins1965}.

Our calculated direct bandgap at the zone center is 2.55~eV, strongly underestimating experiments as expected from density functional theory (DFT). 
In agreement with prior work~\cite{Peelaers2015}, we find that the valence band maximum (VBM) is located on the $I-L$ high-symmetry lines in the Brillouin zone, and yields a slightly smaller indirect bandgap of 2.53~eV. 
A comparison with an earlier work is given in Table~\ref{table2}. 
Our values are consistent with calculations at equivalent level of theory; hybrid functionals slightly overestimates room-temperature experimental bandgaps.

There has been some confusion in the literature about the shape of the Brillouin zone of $\beta$-Ga$_2$O$_3$~\cite{Yamaguchi2004,He2006,Litimein2009,Mohamed2010,Janowitz2011}.
The first band structure using the correct monoclinic variation was reported in 2015~\cite{Peelaers2015}.
It is therefore important to pay close attention when constructing the Brillouin Zone of $\beta$-Ga$_2$O$_3$.

\begin{figure}[h!]
  \centering
  \includegraphics[width=0.99\linewidth]{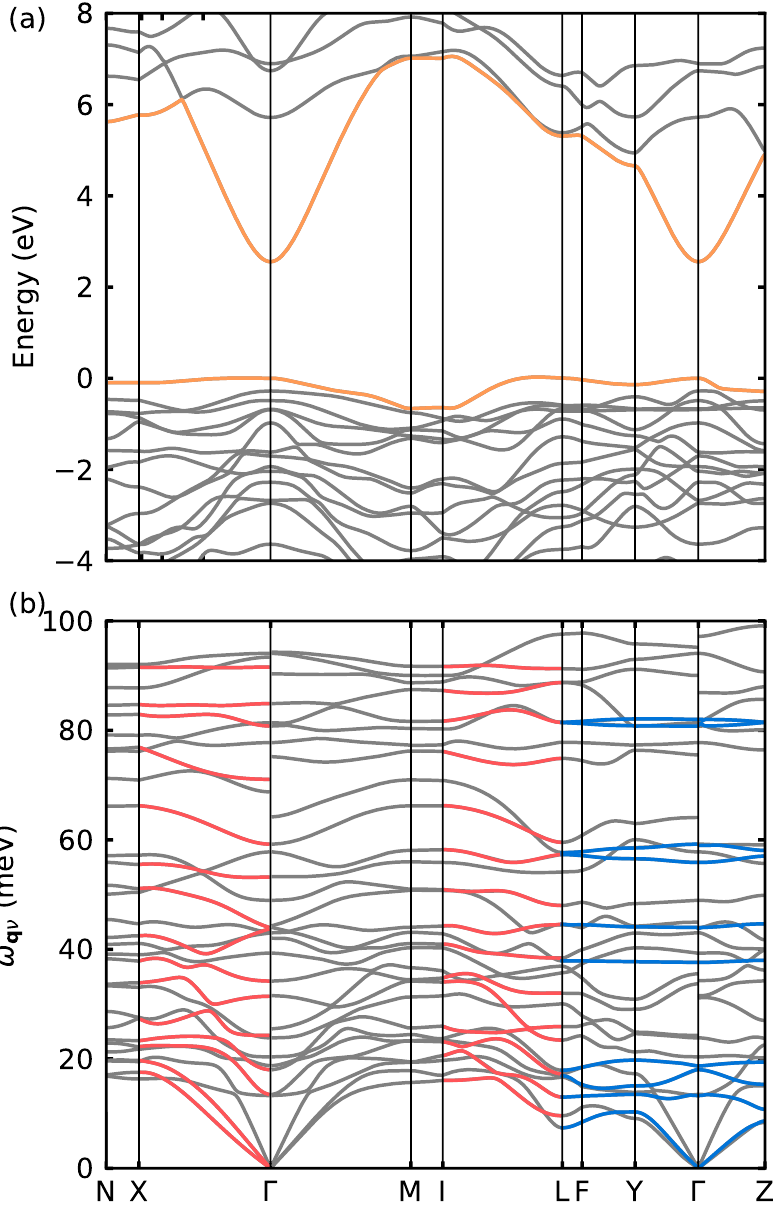}
  \caption{\label{fig2}
  Electronic (a) and phonon (b) bandstructure of $\beta$-Ga$_2$O$_3$ along high-symmetry lines of the monoclinic Brillouin Zone. For the phonons, the gray lines denotes bands with the identity point group while the red lines have a $\pi$ rotation around the [0,1,0] Cartesian axis ($C_2$ point group) and the blue line are lines with inversion symmetry ($C_s(m)$ point group). 
  }
\end{figure}

We note that, as the definition of two of the primitive cell vectors in the Quantum Espresso software are inverted with respect to prior studies, we had to adapt the definition of the high-symmetry points of the Brillouin zone. 
We give the conversion for clarity in Appendix Table~\ref{table3} as well as the value of the four parameters that define the high-symmetry points.
To avoid further confusion, the primitive vectors for the base centered monoclinic Bravais lattice have been modified in Quantum Espresso version 6.5  to use the same definition as in the literature~\cite{Peelaers2015}.
The electronic bandstructure along high-symmetry lines is given in Fig.~\ref{fig2}(a), where the highest valence band and lowest conduction band are highlighted in orange color.

\npdecimalsign{.}
\nprounddigits{2}
\begin{table*}[t!]
  \begin{tabular}{l l n{2}{2} n{2}{2} n{2}{2} n{2}{2} n{2}{2} n{2}{2} n{2}{2} n{2}{2} n{2}{2} n{2}{2} | n{2}{2} n{3}{2} n{2}{2} n{2}{2} n{2}{2}}
  \toprule\\
              &                 &  \multicolumn{10}{c}{Calculated} & \multicolumn{5}{c}{Experiment} \\
Mode symmetry & Activity        &  \multicolumn{4}{c}{This work} & \multicolumn{4}{c}{Mengle~\cite{Mengle2019}} & \multicolumn{2}{c}{Liu~\cite{Liu2007}}  & \multicolumn{2}{c}{Schubert~\cite{Schubert2016}} & \multicolumn{1}{c}{Machon~\cite{Machon2006}} & \multicolumn{2}{c}{Dohy~\cite{Dohy1982}}  \\
              &                 &  & \multicolumn{1}{c}{LO$_X$} & \multicolumn{1}{c}{LO$_Y$} & \multicolumn{1}{c}{LO$_Z$} &  & \multicolumn{1}{c}{LO$_{b_1}$} & \multicolumn{1}{c}{LO$_{b_2}$} & \multicolumn{1}{c}{LO$_{b_3}$} &  &\multicolumn{1}{c}{LO} & & \multicolumn{1}{c}{LO} & \multicolumn{1}{c}{ } & & \multicolumn{1}{c}{LO} \\
\hline 
A$_g$ (1)       & Raman    & 13.3007 & & & & 12.92039  & & & & 12.98115 &  & \multicolumn{1}{c}{-} &  & 13.66306  & 13.76225 & \\ 
B$_g$ (1)       & Raman    & 13.5463 & & & & 13.03198  & & & & 13.89863 &  & \multicolumn{1}{c}{-} &  & 14.08460  & 14.13420 &  \\
B$_g$ (2)       & Raman    & 18.0406 & & & & 17.95663  & & & & 17.51897 &  & \multicolumn{1}{c}{-} &  & 17.94051  & 18.22568 &  \\
A$_u$ (TO1)     & Infrared & 18.7039 & 18.7967 & 18.7130 & 18.7071 & 18.05582  & 18.05830 & 18.10045 & 18.05582 & 17.55616 & 18.16368 & 19.19275 & 19.37873 & \multicolumn{1}{c}{-} & \multicolumn{2}{c}{19.21}  \\
A$_g$ (2)       & Raman    & 20.3690 & & & & 19.86351  & & & & 20.25902 &  & \multicolumn{1}{c}{-} &  & 20.97813 & 20.95333 &   \\
A$_g$ (3)       & Raman    & 23.8311 & & & & 23.11065  & & & & 25.08200 &  & \multicolumn{1}{c}{-} &  & 24.84643 & 24.67285 &   \\ 
B$_u$ (TO1)     & Infrared & 24.2779 & 24.2785 & 24.2801 & 31.1559 & 21.76171  & 22.05555 & 21.76171 & 28.53000 & 23.24704 & 23.61899 & 26.50782 & 33.35175 & \multicolumn{1}{c}{-} & \multicolumn{2}{c}{31.00}  \\
B$_u$ (TO2)     & Infrared & 31.4358 & 31.432  & 33.6936 & 31.5417 & 30.22487  & 32.84589 & 30.22487 & 31.18450 & 31.19442 & 32.79382 & 32.52105 & 35.45948 & \multicolumn{1}{c}{-} & \multicolumn{1}{c}{-} &  \\
B$_u$ (TO3)     & Infrared & 34.2134 & 34.2175 & 35.5251 & 34.2964 & 33.66791  & 34.39569 & 33.66791 & 33.68527 & 32.89301 & 35.16192 & 34.61639 & 37.81518 & \multicolumn{1}{c}{-} & \multicolumn{2}{c}{35.95} \\
A$_u$ (TO2)     & Infrared & 37.6083 & 42.9389 & 37.6085 & 37.6087 & 37.58829  & 38.04455 & 42.53526 & 37.58829 & 36.72412 & 40.35685 & 36.77371 & 42.88613 & \multicolumn{1}{c}{-} & \multicolumn{2}{c}{38.43} \\
A$_g$ (4)       & Raman    & 39.3235 & & & & 38.56776  &          & & & 39.15421 &  & \multicolumn{1}{c}{-} &  & 39.50136 & 39.42697 &   \\
A$_g$ (5)       & Raman    & 43.0353 & & & & 42.66916  &          & & & 42.11743 &  & \multicolumn{1}{c}{-} &  & 42.94812 & 42.89853 &   \\ 
B$_u$ (TO4)     & Infrared & 43.7333 & 43.6516 & 44.0052 & 43.9195 & 43.01508  & 44.48057 & 43.01632 & 43.75154 & 42.60097 & 43.90280 & 44.23756 & 48.22985 & \multicolumn{1}{c}{-} & \multicolumn{2}{c}{46.49} \\
B$_g$ (3)       & Raman    & 44.0051 & & & & 43.45398  &          & & & 43.18369 &  & \multicolumn{1}{c}{-} &  & \multicolumn{1}{c}{-} & 43.76642 &  \\ 
A$_g$ (6)       & Raman    & 48.9601 & & & & 46.64037  &          & & & 52.09816 &  & \multicolumn{1}{c}{-} &  & 51.54023 & 51.45344 &   \\ 
B$_u$ (TO5)     & Infrared & 53.1958 & 53.1945 & 64.0910 & 58.9781 & 52.46267  & 55.20148 & 52.46267 & 55.50276 & 47.54794 & 63.30633 & 53.63556 & 69.67912 & \multicolumn{1}{c}{-} & \multicolumn{2}{c}{56.41}  \\
A$_u$ (TO3)     & Infrared & 55.8780 & 68.7760 & 55.8787  & 55.8788  & 55.56476  & 60.32575 & 67.78960 & 55.56476 & 50.89551 & 60.09514 & 55.63171 & 73.77059 & \multicolumn{1}{c}{-} & \multicolumn{2}{c}{65.09}  \\
A$_g$ (7)       & Raman    & 57.8013 & & & & 56.55663  &          & & & 56.95834 &  & \multicolumn{1}{c}{-} &  & \multicolumn{1}{c}{-} & 58.89249 &  \\ 
B$_g$ (4)       & Raman    & 59.2209 & & & & 58.60485  &          & & & 58.61973 &  & \multicolumn{1}{c}{-} &  & 58.70652  & 58.89249 &  \\
B$_u$ (TO6)     & Infrared & 71.0646 & 71.0662 & 75.5392 & 80.8032 & 70.35855  & 76.45609 & 70.35855 & 82.65406 & 71.20412 & 77.52732 & 70.98095 & 87.90479 & \multicolumn{1}{c}{-} &  \multicolumn{2}{c}{79.35} \\
A$_g$ (8)       & Raman    & 77.7683 & & & & 76.87392  &          & & & 75.27080 &  & \multicolumn{1}{c}{-} &  &  77.94886 & 77.86207 & \\ 
B$_g$ (5)       & Raman    & 80.8029 & & & & 79.81606  &          & & & 77.75049 &  & \multicolumn{1}{c}{-} &  &  80.89969 & 80.71371  & \\
A$_g$ (9)       & Raman    & 81.387  & & & & 79.94873  &          & & & 81.34603 &  & \multicolumn{1}{c}{-} &  &  \multicolumn{1}{c}{-} & 81.45761  &  \\ 
A$_u$ (TO4)     & Infrared & 82.0108 & 93.3977 & 82.0111 & 82.0107 & 80.93316  & 81.22824 & 91.98635 & 80.93316 & 80.32936 & 91.56233 & 82.22632 & 95.46783 & \multicolumn{1}{c}{-} & \multicolumn{2}{c}{82.82}  \\
B$_u$ (TO7)     & Infrared & 84.8999 & 84.8976 & 90.0416 & 86.8960 & 83.81455  & 89.35789 & 83.81455 & 85.16598 & 83.39177 & 90.28529 & 85.84666 & 96.86885 & \multicolumn{1}{c}{-} & \multicolumn{2}{c}{89.27}\\
B$_u$ (TO8)     & Infrared & 91.5781 & 91.5770 & 95.1837 & 97.1831 & 89.96541  & 92.59016 & 89.96541 & 94.75616 & 91.94668 & 94.79831 & 92.18225 & 100.42720 & \multicolumn{1}{c}{-} & \multicolumn{2}{c}{94.23} \\
A$_g$ (10)      & Raman    & 94.0737 & & & & 92.39178  &          & & & 93.94282 &  & \multicolumn{1}{c}{-} &  & 94.71153 & 94.59994 &  \\
  \botrule     
  \end{tabular}
  \caption{\label{table4}
 Phonon frequencies (meV) of $\beta$-Ga$_2$O$_3$ at the zone center using a 16$\times$16$\times$12 $\mathbf{k}$-grid. 
 The three zero-frequency acoustic modes are not reported.
 The infrared experimental values from Dohy \textit{et al.}~\cite{Dohy1982} are measured by transmission and the frequency of the maxima lies between the LO and TO frequencies. 
  }
\end{table*}

We computed the electron effective mass using finite differences, and found 0.267, 0.254 and 0.244 along the $\Gamma$-X, $\Gamma$-Y and $\Gamma$-Z direction, respectively.
The electron effective mass is quite isotropic with an average value of 0.255 as reported in Table~\ref{table2}, which compares well with prior theoretical work, and is also close to the experimental value of 0.28~\cite{Mohamed2010,Janowitz2011}.
This level of agreement gives us confidence that our calculations of electronic transport properties will be reliable.

In contrast, the hole effective mass at the zone-center is highly anisotropic, with very heavy masses along the $\Gamma$-X and $\Gamma$-Y direction, and a small hole mass of 0.35~m$_e$ along the $\Gamma$-Z direction. 
As a result, this should be an ideal hole transport direction.
However, the VBM is not located at the zone centred but 26~meV higher in energy on the I-L line.
The transverse and perpendicular hole effective mass at that point is 3.0~m$_e$ and 3.6~m$_e$, respectively in agreement with previous work~\cite{Yamaguchi2004}. 
As transport properties scale inversely with the effective mass, we expect at least an order of magnitude lower hole mobility than the electron mobility. 

\section{Vibrational properties}\label{sec:vibration}
\subsection{Phonons dispersions}
We now study the vibrational properties of $\beta$-Ga$_2$O$_3$ using density functional perturbation theory (DFPT)~\cite{Gonze1997a,Baroni2001}. 
The calculated phonon bandstructure along the monoclinic Brillouin zone is presented in Fig.~\ref{fig2}(b).

The point groups along high-symmetry lines is either $C_s(m)$ or $C_2$.
The $C_s(m)$ point group contains two symmetry operations: the identity operation $E$ and a mirror plane $\sigma$.
This point group possess two irreducible representations: the phonon branches belonging to the $A'$ irreducible representation are symmetric with respect to both the identity operation and reflection through the mirror plane while the branches belonging to the $A''$ representation are symmetric with respect to the identity but antisymmetric with respect to reflection (coloured in Fig.~\ref{fig2}(b) in gray and blue, respectively). 
The other point group is the $C_2$ point group which contains the identity (gray) and a $\pi$ rotation around the [0,1,0] Cartesian axis (displayed with red lines in Fig.~\ref{fig2}(b)). 
Note that some directions in the Brillouin-Zone are less symmetric and only possess the identity (gray).   
In addition, specific high symmetry points have higher symmetries: 
(i) the point group at the $N$ and $M$ points is $C_i(-1)$ with $A_g$ and $A_u$ symmetry operation; 
(ii) the point group at the $X$, $I$ points is $C_2(2)$ with identity $E$ and $C_2$ with $\pi$ rotation around the [0,1,0] Cartesian axis; 
(iii) the point group at the $\Gamma$, $L$, $Y$, $Z$ point is $C_{2h}(2/m)$ with $A_g$, $B_g$, $A_u$, $B_u$ symmetries. 

\subsection{Infrared and Raman spectra}
The infrared spectrum  as well as polarization and temperature-dependent Raman spectra of bulk $\beta$-Ga$_2$O$_3$ were first measured by Dohy~\textit{et al.}~\cite{Dohy1982} in 1982. 
The measured normal modes frequencies are reported in Table~\ref{table4} along with more recent measurements and previous \textit{ab-initio} values, and are compared to the calculated frequencies from this work. 
Our calculated phonon bandstructure slightly underestimates experiments but are in better agreement than previous calculations.
Overall, our calculations agree with previous theoretical work~\cite{Liu2007,Santia2015,Mengle2019} with a notable difference:
 in agreement with experiments~\cite{Machon2006,Schubert2016} we find that the $A_g(3)$ Raman active mode has a lower frequency than the $B_u$(TO1) mode. 
The highest phonon frequency at the zone center is a LO mode in the $Z$ direction, with a frequency of 97~meV, very close to the experimental value of 100~meV~\cite{Schubert2016}. 
However we note that the highest phonon frequency occurs at the $Z$ point with a value of 99.12~meV (not shown in Table~\ref{table4}).
Our predicted Raman-active phonon frequencies are 2.5\% within the experimental data~\cite{Machon2006} with the largest difference being attributed to the 
$A_g(6)$ mode.
Our predicted infrared-active LO modes are even closer, with deviation of 1.4~\% from experimental data~\cite{Schubert2016}, while the agreement with LO modes is not as good with a deviation of 5.4\%.
%

\subsection{Dielectric constant and Born charges}

The high-frequency dielectric tensor is fairly isotropic, with $\varepsilon_{xx}=3.98$, $\varepsilon_{yy}=4.09$ and $\varepsilon_{zz}=4.08$, slightly overestimating the experimental value of 3.53-3.6~\cite{Passlack1995,Rebien2002,Schmitz1998} obtained as an isotropic average in thin films.
The slight overestimation of the theoretical dielectric tensor is a direct consequence of the underestimation of the bandgap by DFT as the electronic part of the dielectric function is inversely proportional to the bandgap~\cite{Lee2018b}.
We note one experimental work which obtained a direction-dependent dielectric tensor $\varepsilon_{xx}=3.7$, $\varepsilon_{yy}=3.2$ and $\varepsilon_{zz}=3.7$~\cite{Schubert2016} using generalized spectroscopic ellipsometry within the infrared and far-infrared spectral region.
This anisotropy was not observed in another recent experiment reporting $\varepsilon_{xx}=3.6$, $\varepsilon_{yy}=3.58$ and $\varepsilon_{zz}=3.54$~\cite{Sturm2016} also using generalized spectroscopic ellipsometry. 
Our calculations appear to support an isotropic dielectric tensor.
$\beta$-Ga$_2$O$_3$ also possesses one non-zero off-diagonal component of the dielectric tensor, but the computed value was lower than $10^{-4}$ and therefore is not reported.

The computed diagonal Born effective charges are 
Ga$_{\rm I}$ = (2.74, 2.88, 3.04),  
Ga$_{\rm II}$ = (3.23, 3.42, 3.12),
O$_{\rm I}$ = -(1.46, 2.09, 2.47),
O$_{\rm II}$ = -(2.27, 2.25, 1.39),
O$_{\rm III}$ = -(2.22, 1.96, 2.28) in units of electron charge. 
The off-diagonal components are lower than 0.3 and not reported. 

\subsection{Elastic properties}

\begin{table}[t]
  \begin{tabular}{l r r r r r r r r}
  \toprule 
                             & $C_{11}$ & $C_{12}$ & $C_{13}$ &  $C_{15}$ & $C_{22}$ & $C_{23}$ & $C_{25}$ & $C_{33}$ \\
\rule{0pt}{1em}This work     &  GPa     & GPa      &    GPa   &     GPa   &     GPa  &     GPa  &     GPa &   GPa  \\
\hline \\[-0.33cm]                     
LDA                          & 242 & 127 & 140 & -17.7 &  360 & 90.3 & 12.0  & 355 \\
                             &     &$\pm 3.4$ & $\pm 0.0$ & $\pm 0.3$ & & $\pm 0.7$ & $\pm 0.4$ &  \\
\rule{0pt}{1em}Previous      &     &     &     &     &     &   &   & \\
\hline \\[-0.33cm]           
LDA~\cite{Adachi2018}        & 219 & 127 & 169 & -1.4 & 365 & 106 & 3.5 & 344 \\
AM05~\cite{Furthmueller2016} & 223 & 116 & 125 &  -17 & 333 & 75  & 12  & 330 \\
GGA~\cite{Jain2013,Jong2015a}& 199 & 112 & 125 & -2   & 312 & 62  & 1   & 298 \\
PBESOL~\cite{Grashchenko2018}& 227 & 128 & 135 & -3.6 & 335 & 73  & 0   & 313 \\
PBESOL~\cite{Osipov2018}     & 208 & 118 & 146 &    0 & 335 & 83  & 0   & 318 \\
Exp.~\cite{Miller2017}       & 238 & 130 & 152 & -4   & 359 & 78  & 2   & 346 \\
Exp.~\cite{Adachi2018}       & 243 & 128 & 160 & -1.6 & 344 & 71  & 0.4 & 347 \\  
\\  
                              &  $C_{35}$  &  $C_{44}$  & $C_{46}$  & $C_{55}$ & $C_{66}$ & $B_{\rm H}$ & $E_{\rm H}$ & $G_{\rm H}$ \\
\rule{0pt}{1em}This work      &  GPa       &    GPa     &     GPa   &     GPa  &     GPa  & GPa & GPa & GPa  \\
\hline \\[-0.33cm]                                    
LDA                           &   7.7 & 58 & 19.7 & 69 & 97 & 184 & 207 & 79 \\
                              &$\pm 0.5$&  & $\pm 0.3$& &   &     &     & \\
\rule{0pt}{1em}Previous       &       &    &      &    &    &     &     & \\
\hline 
LDA~\cite{Adachi2018}         & 18 & 54 & 13  & 76 & 99  & 189 & 198 & 74 \\
AM05~\cite{Furthmueller2016}  &  7 & 50 & 17  & 69 & 94  & 167 & 194 & 74 \\
GGA~\cite{Jain2013,Jong2015a} & 17 & 39 & 3   & 77 & 95  & 155 & 182 & 70 \\
PBESOL~\cite{Grashchenko2018} & 18 & 45 & 6.4 & 83 & 99  & 177 & 207 & 79 \\
PBESOL~\cite{Osipov2018}      & 19 & 50 & 9   & 77 & 96  & 171 & 192 & 73 \\
Exp.~\cite{Miller2017}        & 19 & 49 & 6   & 91 & 107 & 184 & 213 & 82 \\  
Exp.~\cite{Adachi2018}        &  1 & 48 & 5.6 & 89 & 104 & 183 & 210 & 80 \\ 
\\
                              &$\nu_{\rm H}$ & $A^{\rm U}$ & $v_{\rm B}$ & $v_{\rm P}$ & $v_{\rm G}$ & $v_{\rm av}$ & $\Theta_{\rm D}$ \\
\rule{0pt}{1em}This work      &      &      & km/s & km/s & km/s & km/s & K \\
\hline \\[-0.33cm]                                    
LDA                           & 0.31 & 0.84 & 5.48 & 6.87 & 3.59 & 4.01 & 551 \\
\rule{0pt}{1em}Previous       &      &      &    &      &      & \\
\hline \\[-0.33cm]           
LDA~\cite{Adachi2018}         & 0.33 & 0.93 & 5.55 & 6.86 & 3.49 & 3.91 & 538 \\
AM05~\cite{Furthmueller2016}  & 0.31 & 0.92 & 5.23 & 6.60 & 3.49 & 3.90 & 536 \\
GGA~\cite{Jain2013,Jong2015a} & 0.31 & 1.04 & 5.04 & 6.37 & 3.37 & 3.77 & 518 \\
PBESOL~\cite{Grashchenko2018} & 0.31 & 0.70 & 5.38 & 6.80 & 3.60 & 4.03 & 553 \\
PBESOL~\cite{Osipov2018}      & 0.31 & 0.85 & 5.28 & 6.62 & 3.46 & 3.87 & 532 \\
Exp.~\cite{Miller2017}        & 0.31 & 0.90 & 5.48 & 6.91 & 3.65 & 4.08 & 561 \\ 
Exp.~\cite{Adachi2018}        & 0.31 & 0.88 & 5.47 & 6.88 & 3.62 & 4.04 & 556 \\ 
  \botrule 
  \end{tabular}
  \caption{\label{table5}
  Comparison between our calculated elastic constants $C_{ij}$, bulk modulus $B_{\rm H}$, Young modulus $E_{\rm H}$, shear modulus $G_{\rm H}$, Poisson's ratio $\nu_{\rm H}$, universal elastic anisotropy $A^{\rm U}$, bulk sound velocity $v_{\rm B}$, compressional velocity $v_{\rm P}$, shear velocity $v_{\rm G}$, average velocity $v_{\rm av}$, and Debye temperature $\Theta_{\rm D}$, and prior theoretical and experimental work. The subscript H denotes the Void-Reuss-Hill averaging approximation. 
  }
\end{table}

\begin{figure*}[ht]
  \centering
  \includegraphics[width=0.89\linewidth]{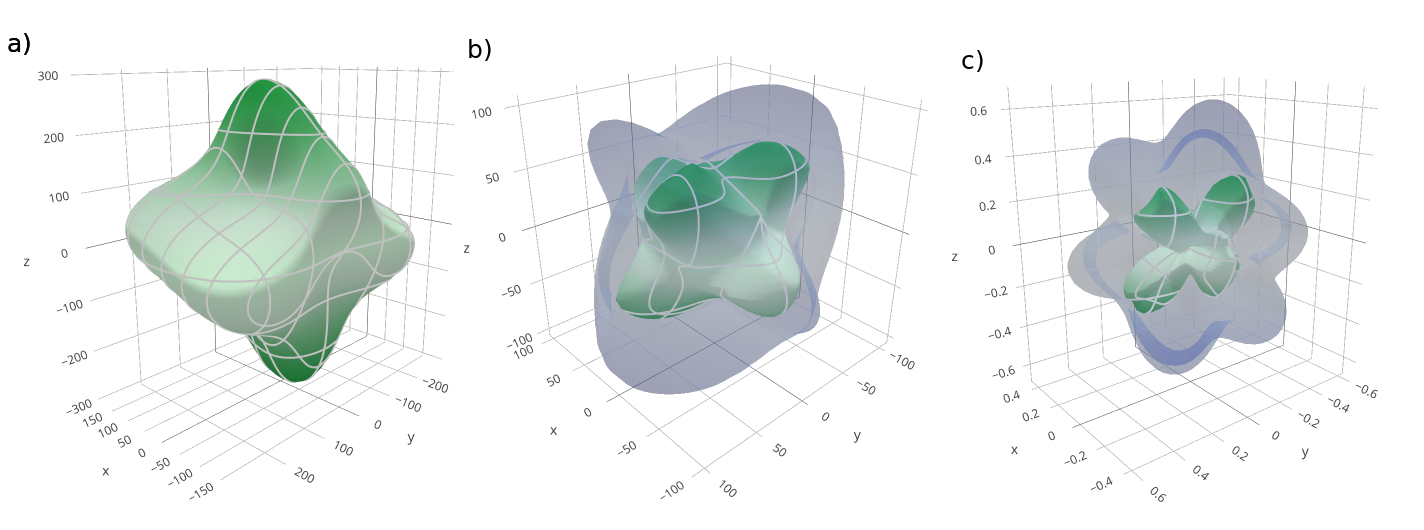}
  \caption{\label{fig3}
   Spatial dependence of the (a) Young modulus, (b) shear modulus and (c) Poisson's ratio  of $\beta$-Ga$_2$O$_3$ using the ELATE software~\cite{Gaillac2016} for visualization of second-order elastic constants. In (b,c) the blue and green surfaces represents the maximum and minimum of the third angle parametrization, see text. The directions $x$, $y$, and $z$ represents the increments along the $a$, $b$ and $c$ directions of the primitive cell shown in Fig.~\ref{fig1}.
  }
\end{figure*}

The stiffness $C_{ij}$ and compliance $S_{ij} = C_{ij}^{-1}$ tensors link the stress tensor to the strain tensor following the generalized Hooke's law:
\begin{align}
\sigma_{ij}      =& C_{ijkl} \varepsilon_{kl} \\
\varepsilon_{ij} =& S_{ijkl} \sigma_{kl},
\end{align}
where Einstein's notation is implied. 

The Young modulus $E$ is the linear response of a material to a uniaxial stress where the response is measured in the direction of the applied stress and the Bulk modulus $B$ is the response to an isotropic stress.  
The Young and bulk moduli can therefore be expressed as a function of a single unit vector in Cartesian space expressed in spherical coordinates $0\leq \theta \leq \pi$ and $0\leq \phi \leq 2\pi$ as $\mathbf{u} = (\sin \theta \cos \phi, \sin \theta \sin \phi, \cos \theta)$~\cite{Marmier2010}:
\begin{align}\label{eq:youngmodulus}
 E(\theta,\phi) =& \frac{1}{u_i u_j u_k u_l S_{ijkl}} \\
 B(\theta,\phi) =& \frac{1}{u_i u_j S_{ijkk}},
\end{align}
where in the case of the Young modulus we have transformed the head of the compliance tensor from the Cartesian basis to a new basis whose first unit vector is $\mathbf{u}$ following the transformation:
\begin{equation}
S_{1111}^{'} = a_{1i} a_{1j} a_{1k} a_{1l} S_{ijkl} = u_i u_j u_k u_l S_{ijkl},
\end{equation}  
where $a_{ij}$ are the direction cosine specifying the angle between the i$^{th}$ axis of the new basis and the j$^{th}$ axis of the initial basis. 
The Bulk modulus is simpler because it is obtained by applying an isotropic stress (pressure $p$) such that $\varepsilon_{ij}=-pS_{ijkk}$.

Other elastic properties such as the shear modulus $G$ or Poisson's ratio $\nu$ depend on the direction in which the stress is applied $\mathbf{u}$ but also the orthogonal direction in which the response is measured $\mathbf{v}$ and can be parametrized with three angles $\theta$, $\phi$ and $0\leq \xi \leq 2\pi$: 
\begin{equation}
\mathbf{v} = \begin{bmatrix}
\cos\theta \cos \phi \cos \xi - \sin \phi \sin \xi \\
\cos\theta \sin \phi \cos \xi  + \cos \phi \sin \xi \\
- \sin \theta \cos \xi  
\end{bmatrix}.
\end{equation}

The shear modulus and Poisson's ratio can therefore be obtained as:
\begin{align}\label{eq:shearmodulus}
 G(\theta,\phi,\xi) =& \frac{1}{4 u_i v_j u_k v_l S_{ijkl}} \\
 \nu(\theta,\phi,\xi) =& -\frac{u_i u_j v_k v_l S_{ijkl}}{u_i u_j u_k u_l S_{ijkl}}. \label{eq:poissonratio}
\end{align}
We note that for the elastic properties studied here we only need up to two vectors (or three angles) in the new basis because the direction of applied stress and measured response are orthogonal but a general elastic property where this was not the case would require three vectors (or four angles) in the transformed basis.

These elastic properties can be averaged by direct integration on the unit sphere to give the standard Young modulus, bulk modulus, shear modulus and Poisson's ratio.  
However, very popular averaging approximations have been developed including the Voigt approximation where the average bulk and shear moduli are given by~\cite{Hill1952}: 
\begin{align}\label{eq:Bv}
9B_{\rm V}     =& \, C_{11} + C_{22} + C_{33} + 2(C_{12}+C_{13}+C_{23}),\\
15 G_{\rm V}   =& \, C_{11} + C_{22} + C_{33} - (C_{12}+C_{13}+C_{23})   \nonumber \\
               +& \, 3 (C_{44} + C_{55} + C_{66}). \label{eq:Br}
\end{align}

In the Reuss approximation, the bulk and shear modulus are defined as~\cite{Hill1952}: 
\begin{align}
B_{\rm R}^{-1}    =& \, S_{11}+S_{22}+S_{33} + 2(S_{12}+S_{13}+S_{23}), \label{eq:Gv} \\
15 G_{\rm R}^{-1} =& \, 4(S_{11}+S_{22}+ S_{33}) - 4(S_{12}+S_{13}+S_{23}) \nonumber \\
                  +& \, 3(S_{44} + S_{55} + S_{66}).\label{eq:Gr}
\end{align}
The Voigt approximation provides an upper bound for the bulk and shear moduli, while the Reuss approximation gives a lower bound. 
We can therefore define the arithmetic mean, referred to as the Void-Reuss-Hill approximation~\cite{Hill1952}, as $B_{\rm H}=(B_{\rm V}+B_{\rm R})/2$ and 
$G_{\rm H}=(G_{\rm V}+G_{\rm R})/2$. 
We then express the effective Young $E$ modulus and Poisson ratio $\nu$ as:
\begin{align}
E   =& \, 9B G/(3B+G) \label{eq:E}, \\
\nu =& \,  (3B-2G)/(6B+2G), \label{eq:nu}
\end{align}
where the relations apply to the Voigt, Reuss and Hill approximation of the Young, bulk and shear moduli and the Poisson's ratio. 
We can also define the universal elastic anisotropy as~\cite{Jong2015a}:
\begin{equation}\label{eq:Au}
A^{\rm U} = 5(G_{\rm V}/G_{\rm R}) + (B_{\rm V}/B_{\rm R}) -6.
\end{equation}
%
Finally, we can obtain the bulk sound velocity $v_{\rm B}$, the compressional velocity $v_{\rm P}$, shear velocity $v_{\rm G}$ and the average sound velocity $v_{\rm av}$ as:
\begin{align}
v_{\rm B}  &= \sqrt{B/\rho}  \\ \label{eq:vB}
v_{\rm P}  &= \sqrt{\Big(B+\frac{4}{3}G\Big)\frac{1}{\rho}}\\
v_{\rm G}  &= \sqrt{G/\rho} \\
v_{\rm av} &= \bigg[\frac{1}{3} \Big( \frac{2}{v_{\rm G}^3} + \frac{1}{v_{\rm P}^3}\Big)\bigg]^{-\frac{1}{3}},  \label{eq:vag}  
\end{align}
where $\rho$ is the average mass density.
%
Using the average sound velocity, the Debye temperature can be estimated within the Debye model as:
\begin{equation}\label{eq:D}
\Theta_{\rm D} = \frac{h}{k_B} v_{\rm av} \Big[ \frac{3 N_{at}}{4 \pi \rho }\Big]^{1/3},
\end{equation}
where $h$, $k_B$ and $N_{at}$ are the Planck constant, Boltzmann constant and the number of atoms in the primitive cell, respectively.

We studied the elastic properties of $\beta$-Ga$_2$O$_3$ using the thermo\_pw code~\cite{Corso2016}.
The stiffness tensor of Laue class $C_{2h}$ for base centered monoclinic crystals has 13 independent elastic constants written in Voigt notation as follow: $C_{11}$, $C_{12}$, $C_{13}$, $C_{15}$, $C_{22}$, $C_{23}$, $C_{25}$, $C_{33}$, $C_{35}$, $C_{44}$, $C_{46}$, $C_{55}$ and, $C_{66}$. 
The stiffness matrix $C_{ij}$ was obtained by third-order polynomial fitting using 12 deformations  with strain intervals of 0.001 to remain in the linear regime. 
The strains were applied along the crystal lattice vector of the $\beta$-Ga$_2$O$_3$ primitive cell presented in Fig.~\ref{fig1} such that the resulting stiffness matrix is expressed in that basis.  
For each strain, the ions were relaxed to their equilibrium positions with a very tight convergence threshold of 4$\cdot$10$^{-6}$~Ry/\AA\,on forces.
We used a 160~Ry energy cutoff on planewaves and a 12$\times$12$\times$9 $\mathbf{k}$-point grid.
All the elastic coefficients and elastic properties are reported in Table~\ref{table5}.
Our calculations compare well with prior theoretical work and with
resonant ultrasound spectroscopy coupled with laser-Doppler interferometry~\cite{Adachi2018}.
We computed all coefficients independently such that we can estimate off-diagonal accuracy when symmetry constrain are not precisely fulfilled. 
The most sensitive coefficient is the $C_{12}$, with an accuracy of $\pm 3.4$~GPa.
%
%

Using Eqs.~\eqref{eq:Bv}-\eqref{eq:nu}, we obtained a bulk modulus of 184~GPa, a Young modulus of 207~GPa, a shear modulus of 79~GPa and a Poisson's ratio of 0.313.
Those numbers agree well with recent experimental elastic constants of $B=183$~GPa, $E=210$~GPa, $G=80$~GPa and $\nu=0.31$~\cite{Adachi2018}. 
Finally, using Eqs.~\eqref{eq:Au}, \eqref{eq:vB}-\eqref{eq:vag} and \eqref{eq:D} we compute the universal elastic anisotropy $A^{\rm U}$ to be 0.84, the average sound velocity to be 4.01~km/s and the estimated Debye temperature $\Theta_{\rm D}$ to be 551~K.

Using the ELATE software~\cite{Marmier2010,Gaillac2016}, we show in Fig.~\ref{fig3}(a) the parametrized Young modulus of Eq.~\eqref{eq:youngmodulus} as a parametrized three dimensional surface and in Fig.~\ref{fig3}(b,c) 
the parametrized shear modulus and Poisson's ratio of Eqs.~\eqref{eq:shearmodulus} and \eqref{eq:poissonratio} where the maximum and minimum value of the third angle is shown in blue and green, respectively.   
Compared to simple semiconductors where the bulk modulus is spherical, $\beta$-Ga$_2$O$_3$ is strongly anisotropic. For example the Young modulus has a minimum value of 134~GPa
in the $xz$ plane with a unit vector (0.94, 0, 0.34) while the maximum value of the Young modulus is 293~GPa in the (0.34, 0.93, 0.13) direction.
In the case of the shear modulus and the Poisson's ratio presented in Fig.~\ref{fig3}, they are also highly anisotropic with values ranging from 50~GPa to 133~GPa for the shear modulus and from 0 to 0.67 for the Poisson's ratio which displays a flower-like shape along the diagonal axes.

%
%
%
%
%

\section{Carrier mobility}\label{sec:mob}

We now analyze the intrinsic carrier transport properties of $\beta$-Ga$_2$O$_3$. 
We compute the \textit{ab-initio} drift carrier mobility 
\begin{equation}\label{eq:btemob}
\mu_{\alpha\beta} = \frac{e}{V_{\rm uc}n_{\rm c}}\sum_n \int \frac{\mathrm{d}^3k}{\Omega_{\rm BZ}} v_{n\mathbf{k}}^{\alpha} \partial_{E_\beta} f_{n\mathbf{k}}
\end{equation}
 through the linear response $\partial_{E_{\beta}} f_{n\mathbf{k}}$ of the electronic occupation function $f_{n\mathbf{k}}$ to the electric field $\mathbf{E}$ and where $V_{\rm uc}$ is the unit cell volume, $\Omega_{\rm BZ}$ the first Brillouin zone volume and, $n_{\rm c} = (1/V_{\rm uc})\sum_n \int (\mathrm{d}^3 k/ \Omega_{\rm BZ}) f_{n\mathbf{k}} $ is the carrier concentration. We solve the linearized Boltzmann transport equation (BTE)~\cite{Ponce2018,Macheda2018,Ponce2020}:
\begin{align}\label{eq:iterwithbimpl}
 \partial_{E_{\beta}} f_{n\mathbf{k}} =& e v_{n\mathbf{k}}^\beta \frac{\partial f_{n\mathbf{k}}}{\partial \varepsilon_{n\mathbf{k}}} \tau_{n\mathbf{k}} + \frac{2\pi\tau_{n\mathbf{k}}}{\hbar}
  \sum_{m\nu} \!\int\! \frac{\mathrm{d}^3 q}{\Omega_{\mathrm{BZ}}} | g_{mn\nu}(\mathbf{k},\mathbf{q})|^2 \nonumber \\ 
 \times & \Big[(n_{\mathbf{q}\nu}+1-f_{n\mathbf{k}})\delta(\varepsilon_{n\mathbf{k}} - \varepsilon_{m\mathbf{k+q}}  + \hbar \omega_{\mathbf{q}\nu} ) \nonumber  \\
  +  (n_{\mathbf{q} \nu} +& f_{n\mathbf{k}})\delta(\varepsilon_{n\mathbf{k}} - \varepsilon_{m\mathbf{k+q}}  - \hbar \omega_{\mathbf{q}\nu}) \Big] \partial_{E_{\beta}} f_{m\mathbf{k}+\mathbf{q}},
\end{align}
with $\tau_{n\mathbf{k}}$ being the total scattering lifetime:
\begin{align}\label{eq:scattering_rate}
  \tau_{n\mathbf{k}}^{-1} =& \frac{2\pi}{\hbar} \sum_{m\nu} \!\int\! \frac{d\mathbf{q}}{\Omega_{\text{BZ}}} | g_{mn\nu}(\mathbf{k,q})|^2 \nonumber \\
  \times & \big[ (n_{\mathbf{q}\nu} +1 - f_{m\mathbf{k+q}} ) \delta( \varepsilon_{n\mathbf{k}} - \varepsilon_{m\mathbf{k+q}}   -  \hbar \omega_{\mathbf{q}\nu}) \nonumber \\
   + &  (n_{\mathbf{q}\nu}  +   f_{m\mathbf{k+q}} )\delta(\varepsilon_{n\mathbf{k}} - \varepsilon_{m\mathbf{k+q}}  +  \hbar \omega_{\mathbf{q}\nu}) \big]. 
\end{align}
Here $v_{n\mathbf{k}}$ is the electronic velocity of the eigenstates $\varepsilon_{n\mathbf{k}}$, $f_{n\mathbf{k}}$ is the Fermi-Dirac occupation, and $n_{\mathbf{q}\nu}$ is the Bose-Einstein distribution function. 
The electron-phonon matrix elements $g_{mn\nu}(\mathbf{k,q})$ are the probability amplitude for scattering from an initial state $n\mathbf{k}$ to a final state $m\mathbf{k+q}$ via the emission or absorption of a phonon of frequency $\omega_{\mathbf{q}\nu}$.
A common approximation known as the self-energy relaxation time approximation (SERTA) consists in neglecting the second term on the right-hand side of Eq.~\eqref{eq:iterwithbimpl}. The mobility then takes the simpler form:
\begin{equation}
\mu_{\alpha\beta}^{\rm SERTA} = \frac{1}{V_{\rm uc} n_{\rm c}}\sum_n \int \frac{\mathrm{d}^3 k}{\Omega_{\rm BZ}} v_{n\mathbf{k}}^{\alpha} v_{n\mathbf{k}}^{\beta} \tau_{n\mathbf{k}}.
\end{equation} 
We used the EPW software~\cite{Giustino2007,Ponce2016a} to interpolate the electron-phonon matrix element $g_{mn\nu}(\mathbf{k,q})$ from a coarse $8 \times 8 \times 6$ $\mathbf{k}$-points and 
$4 \times 4 \times 3$ $\mathbf{q}$-points grids to a dense  $160 \times 160 \times 120$ $\mathbf{k}$ and $\mathbf{q}$ grids, as required to converge the electron mobility. The interpolation uses maximally localized Wannier function~\cite{Marzari2012} and the Wannier90 software~\cite{Pizzi2020}.
We used 22 Wannier functions of initial $s$ character centered on the gallium atoms and of $p$ character centered on the oxygen atoms.
The Dirac delta function in Eqs.~\eqref{eq:iterwithbimpl} and \eqref{eq:scattering_rate} were computed
using the adaptive smearing method of Refs.~\cite{Li2014b,Li2015,Macheda2018}.  

\begin{figure}[t]
  \centering
  \includegraphics[width=0.99\linewidth]{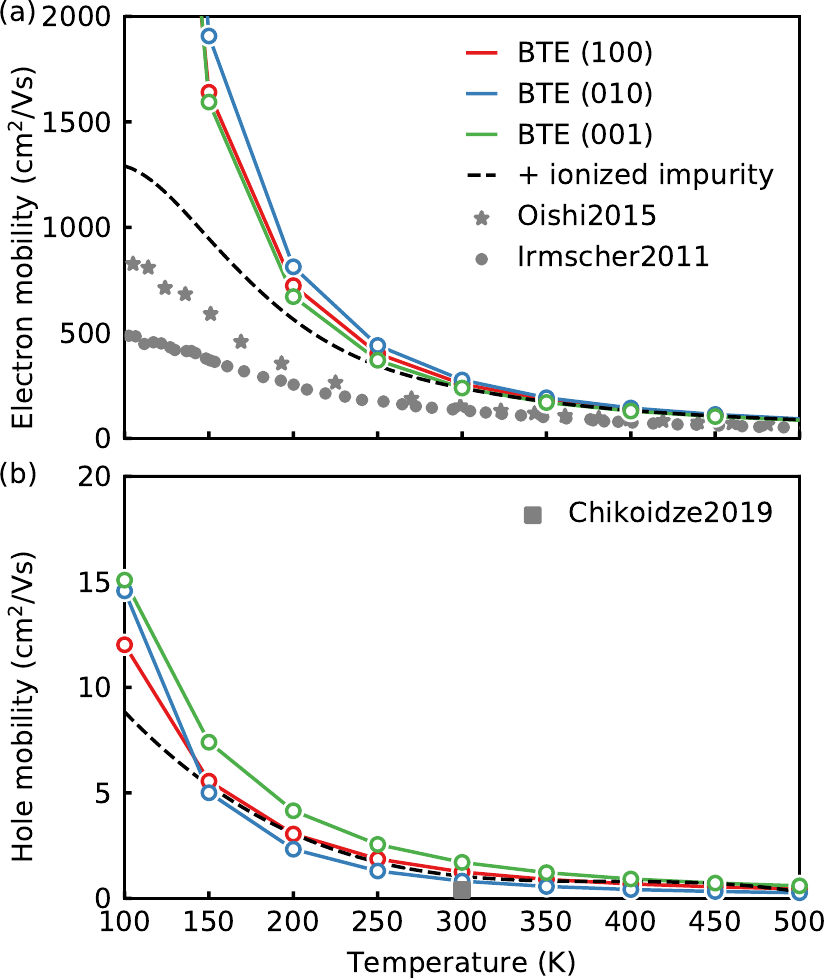}
  \caption{\label{fig4}
   (a) Electron and (b) hole drift mobility of $\beta$-Ga$_2$O$_3$ using the Boltzmann transport equation along the three principal directions where the dashed line indicate the direction-averaged drift mobility including the effect of $10^{15}$~cm$^{-3}$ ionized impurity scattering.
   The experimental data of Oishi~\textit{et al.}~\cite{Oishi2015}, Irmscher~\textit{et al.}~\cite{Irmscher2011} and   Chikoidze~\textit{et al.}~\cite{Chikoidze2019} are Hall measurement of the mobility.    
  }
\end{figure}

To reduce computational cost, we computed separately the electron and hole mobility by explicitly interpolating only the matrix elements for which their 
electronic eigenvalues at $\mathbf{k}$ and $\mathbf{k+q}$ where within 0.3~eV of the band edges. 
We also relied on crystal symmetries to decrease the number of $\mathbf{k}$-points. 
In the case of the electron mobility we explicitly interpolated 13,516 $\mathbf{k}$ and 101,346 $\mathbf{q}$-points, instead of the 3,072,000 points that would have been required by computing all the points from the  $160 \times 160 \times 120$ grid.
In the case of the hole mobility, owing to very flat bands the majority of grid points 
contribute to the hole mobility, as can be seen in Fig.~\ref{fig2}(a). 
Thus the computational cost is much higher and our densest interpolated grid is 56$\times$56$\times$42 points, which corresponds to 55,892  $\mathbf{k}$-points and 131,712 $\mathbf{q}$-points explicitly computed.

\begin{figure*}[ht!]
  \centering
  \includegraphics[width=0.99\linewidth]{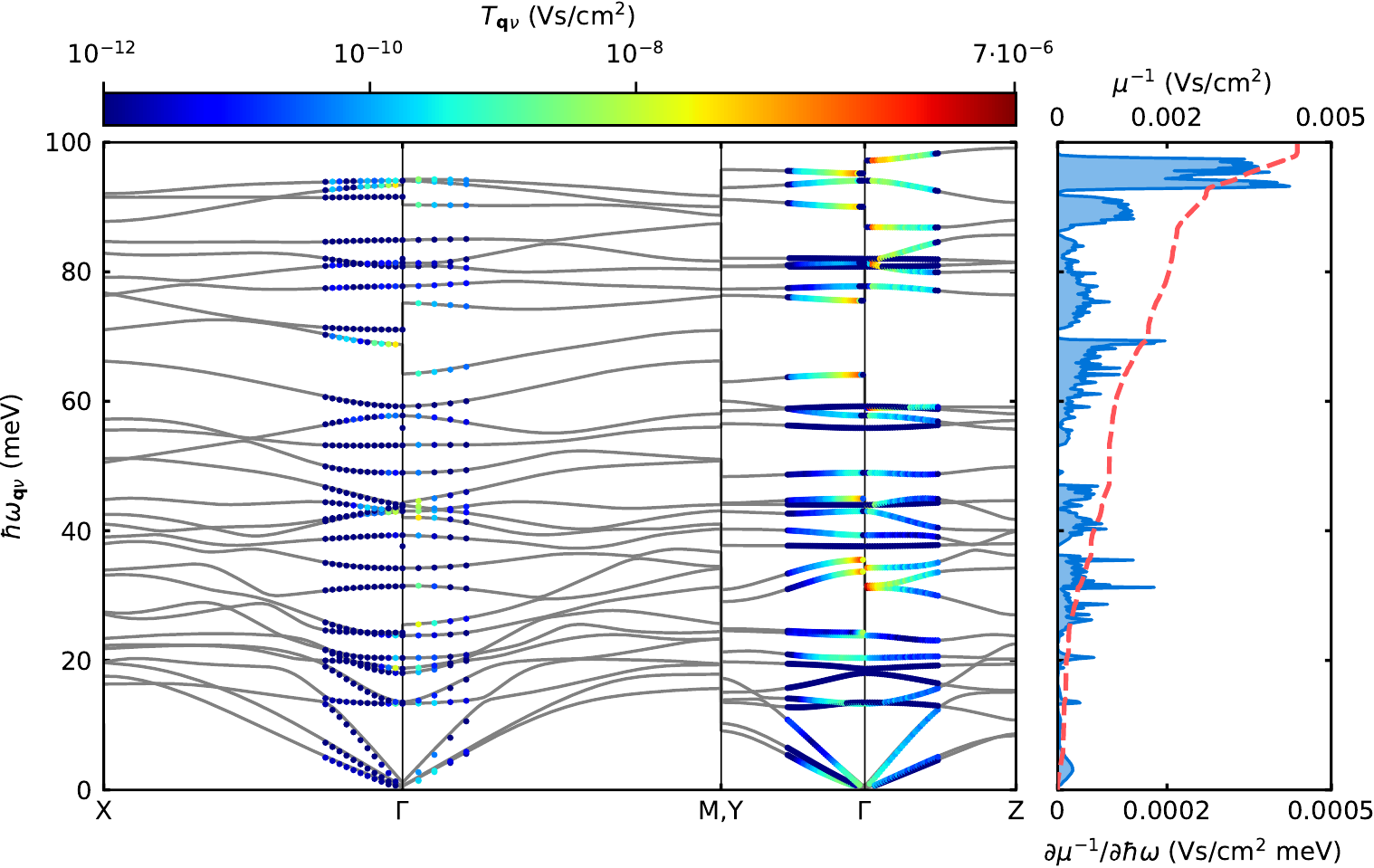}
  \caption{\label{fig5}
   Direction-averaged momentum and mode-resolved contribution to the inverse electron mobility $T_{\mathbf{q}\nu}$ at room temperature (left) as well as spectral decomposition of the inverse mobility (right). The dashed red line represent the cumulative integral of each spectrum $\partial \mu^{-1}/ \partial \hbar \omega $ and adds up to the inverse electron mobility in the self-energy relaxation time approximation.      
  }
\end{figure*}

We obtained the following room-temperature electron and hole drift mobility tensor (cm$^2$/Vs) in the SERTA:
\begin{equation}
\!\!\!\!\!\!\mu_{\alpha\beta,e}^{\rm SERTA} \!\!=\!\! \begin{bmatrix}
170 & 0 & 2.6\\
0 & 165 & 0 \\
2.6 & 0 & 166  
\end{bmatrix}\!,\!
 \mu_{\alpha\beta,h}^{\rm SERTA} \!\!=\!\! \begin{bmatrix}
\phantom{-}1.1 & 0 & -0.3\\
\phantom{-}0 & 0.6 & \phantom{-}0 \\
-0.3 & 0 & \phantom{-}1.6
\end{bmatrix}.
\end{equation}
The result using the self-consistent BTE are
\begin{equation}
\!\! \mu_{\alpha\beta,e}^{\rm BTE} \!=\! \begin{bmatrix}
258 & 0 & 7.5\\
0 & 277 & 0 \\
7.5 & 0 & 239  
\end{bmatrix}\!,\!
 \mu_{\alpha\beta,h}^{\rm BTE} \!=\! \begin{bmatrix}
\phantom{-}1.2 & 0 & -0.2\\
\phantom{-}0 & 0.8 & \phantom{-}0 \\
-0.2 & 0 & \phantom{-}1.7
\end{bmatrix}.
\end{equation}
Interestingly, although the electron effective mass is isotropic (see Table~\ref{table2}), we observe about 15\% anisotropy for the electron mobility resulting from anisotropic electron-phonon scattering.
This result is in line with the recently observed 10-15\% anisotropy in the electron mobility of $\beta$-Ga$_2$O$_3$~\cite{Wong2016}. 
Based on our convergence study with increasing fine grid size, we estimate an accuracy of $\pm$3~cm$^2$/Vs for the electron mobility and $\pm$0.5~cm$^2$/Vs on the hole mobility. 
The anisotropy of the hole mobility is within the uncertainty of the calculations.

The temperature dependence of the BTE electron and hole mobility as a function of temperature is presented in Fig.~\ref{fig4}, 
slightly overestimating experimental data. 
The isotropic average of the electron and hole mobility are 258~cm$^2$/Vs and 1.2~cm$^2$/Vs, respectively.  
To our knowledge, this is the first time that the hole mobility of $\beta$-Ga$_2$O$_3$ is computed from first-principles.

Our room-temperature value of the electron mobility of $\beta$-Ga$_2$O$_3$ is slightly higher than prior theoretical studies: Ref.~\cite{Ghosh2016} gives 115~cm$^2$/Vs at a carrier concentration of $10^{17}$~cm$^{-3}$ using Rode's method~\cite{Rode1975}, and 200~cm$^2$/Vs using $\mathbf{k}\cdot\mathbf{p}$ perturbation theory~\cite{Ma2016}.
Ref.~\cite{Kang2017} obtained an electron mobility of 155~cm$^2$/Vs using the SERTA, in close agreement with our SERTA value of 167~cm$^2$/Vs.

The overestimation with respect to experimental electron mobility can be traced back to the fact that our calculated electron effective mass is 7\% smaller than in experiments and that electron-phonon matrix elements are dominated by Fr\"ohlich polar scattering, which in turn scales with the dielectric constant. 
Our calculated dielectric constant is approximately 11\% higher than in experiments. 
Taken together, these estimates indicate that our calculation underestimate the Fr\"ohlich coupling by approximately 13\%. 
In Ref.~\cite{Ponce2019} we have shown that the mobility is inversely proportional to the Fr\"ohlich coupling and effective mass, therefore
we expect that the use of DFT leads to an overestimation of the mobility by approximately 24\%.
Experimental Hall electron mobilities of 125~cm$^2$/Vs~\cite{Irmscher2011} and 152~cm$^2$/Vs~\cite{Oishi2015} were reported and are consistent with our findings.

Since lattice scattering becomes negligible at low temperature, the mobility computed using Eq.~\eqref{eq:btemob} diverges when $T$ tends to zero. 
At low temperature other scattering mechanisms dominate carrier transport including defect~\cite{Lu2020} and impurity scattering~\cite{Ponce2018}. 
The impurity scattering may be included using the semi-empirical model developed by Brooks and Herring~\cite{Brooks1951,Li1977}.
The ionized-impurity limited mobility $\mu_{\rm i}$ can be evaluated analytically assuming spherical energy surfaces, negligible electron-electron interactions, and complete ionization of the impurities: 
  \begin{equation}\label{BHequation}
  \mu_{\rm i} = \frac{2^{7/2} \epsilon_s^2 (k_{\rm B}T)^{3/2} }{\pi^{3/2} e^3 \sqrt{m_d^*} \,n_{\rm i} 
  G(b)} \quad \bigg[\frac{\rm{cm^2}}{\rm{Vs}}\bigg],
  \end{equation}
where $G(b) = \ln(b+1) - b/(b+1)$, $b = 24\pi m_d^* \epsilon_s (k_{\rm B}T )^{2}/e^2 h^2 n'$, and
$n'=n(2-n_{\rm h}/n_{\rm i})$.
Here $m_d^*=0.26\,m_0$ and $3.39\,m_0$ is the density-of-state effective mass for the electron and hole, respectively, $n$ and $n_{\rm i}$ are the electron or hole densities and 
the density of ionized impurities, respectively, $\epsilon_s=4.05\epsilon_0$ is the average dielectric constant, $\epsilon_0$ is the permittivity of vacuum, and $h$ is Planck's constant. 
In the above expressions, the concentrations are expressed in cm$^{-3}$, and the temperature $T$ is in K.
The mobility including phonon ($\mu$) and impurity ($\mu_i$) scattering can be computed using the 
 mixed-scattering formula~\cite{Li1977} $\mu_l \big[ 1 + X^2\{ \text{ci}(X)\cos(X) + \sin(X)(\text{si}(X)-\frac{\pi}{2})\}  \big]$, where $X^2 = 6\mu/\mu_i$ and ci(X) and si(X) are the cosine and sine integrals. 
The resulting combined mobility for a concentration of $10^{15}$~cm$^{-3}$ of ionized impurity is shown with a dashed line in Fig.~\ref{fig4}, improving the agreement with experiment in the low temperature regime.

Finally, to shed light on the microscopical mechanisms driving the electron mobility in $\beta$-Ga$_2$O$_3$
we computed the isotropic average of the momentum and mode resolved contribution to the SERTA mobility as 
\begin{equation}
\mu = \sum_{\mathbf{q}\nu} T_{\mathbf{q}\nu}^{-1},
\end{equation}
where the mode resolved inverse mobility $T_{\mathbf{q}\nu}$ is
\begin{align}
T_{\mathbf{q}\nu} =& \frac{6\pi}{\hbar} V_{\rm uc} n_{\rm c} \sum_{mn,\alpha} \int d^3 k \frac{ w_{\mathbf{q}} | g_{mn\nu}(\mathbf{k,q})|^2}{v_{n\mathbf{k}}^{\alpha} v_{n\mathbf{k}}^{\alpha}} \nonumber \\
  \times & \big[ (n_{\mathbf{q}\nu} +1 - f_{m\mathbf{k+q}} ) \delta( \varepsilon_{n\mathbf{k}} - \varepsilon_{m\mathbf{k+q}}   -  \hbar \omega_{\mathbf{q}\nu}) \nonumber \\
   + &  (n_{\mathbf{q}\nu}  +   f_{m\mathbf{k+q}} )\delta(\varepsilon_{n\mathbf{k}} - \varepsilon_{m\mathbf{k+q}}  +  \hbar \omega_{\mathbf{q}\nu}) \big],
\end{align}
where $w_{\mathbf{q}}$ is the weight of the $\mathbf{q}$-point.

We show in Fig.~\ref{fig5} the mode contribution to the inverse mobility as well as the density of state inverse mobility along with the cumulative integral (dashed red line). 
The mode contribution spans a region close to the zone center, since as discussed above, larger momenta have negligible contribution to the mobility. 
The spectral decomposition is separated into three defined energy regions: low energy ($\hbar\omega < 50$~meV), middle energy (50~meV $\leq \hbar\omega < 71$~meV), and high energy ($\hbar\omega \geq 71$~meV) regions.
The high energy phonons alone account for 62\% of the inverse mobility at room temperature followed by the low energy phonons (22\%) and middle energy phonons (16\%).  
We mention the following 10 modes, in relation with Table~\ref{table4}, that contributes significantly to reducing the mobility: the B$_u$ (LO$_z$1-3,8) and B$_u$ (LO$_y$2-3, 5-8) modes. 
Interestingly, all the dominant modes have B$_u$ symmetry and are longitudinal optical modes.

As can be seen on the left-side of Fig.~\ref{fig5}, the spectral decomposition of the mode contribution to the inverse mobility is complex, with many modes contributing to the mobility.  
Such complexity in the phonon spectrum of $\beta$-Ga$_2$O$_3$ with 30 crossing and intertwined phonon branches translates into many ways for the electrons to interact with the bosonic continuum yielding increased scattering and reduced mobility. 
It is worth comparing such behavior of the electron scattering with a related material, wurtzite GaN that possess similar electron effective mass $\approx$ 0.2-0.3~m$_e$. 
In the nitride compound, the phonon bandstructure is composed of 12 modes clearly separated by a 20~meV gap~\cite{Ponce2019b}.
This translates into a reduced scattering with two dominant scattering at around 2~meV and 92~meV~\cite{Ponce2019c} and explains why the electron mobility in wurzite GaN is four times larger than in $\beta$-Ga$_2$O$_3$ despites similar effective masses.

\section{Baliga's figure of merit}\label{sec:fom}

Figures of merit have been introduced as a way to quantify the influence of materials parameters on the performance of semiconductor devices. 
The most common figures of merit include the Johnson figure of merit (JFOM) which assess the quality of a semiconductor for high frequency power transistor application~\cite{Johnson1965}, the Keyes' figure of merit (KFOM) which quantifies the thermal limitation of transistors switching frequency~\cite{Keyes1972} and the Baliga's figure of merit (BFOM)~\cite{Baliga1982}. 
In this work we focus solely on the BFOM which is used to identify material's parameters so as to minimize losses in power field effect transistors~\cite{Galazka2018}.
The BFOM relies upon the assumption that power losses are solely due to power dissipation in the on-state by current flow through the on-resistance of the device.  
As a result, the BFOM is used for device operating at low frequency where the conduction losses are dominant.

\begin{figure*}[t]
  \centering
  \includegraphics[width=0.95\linewidth]{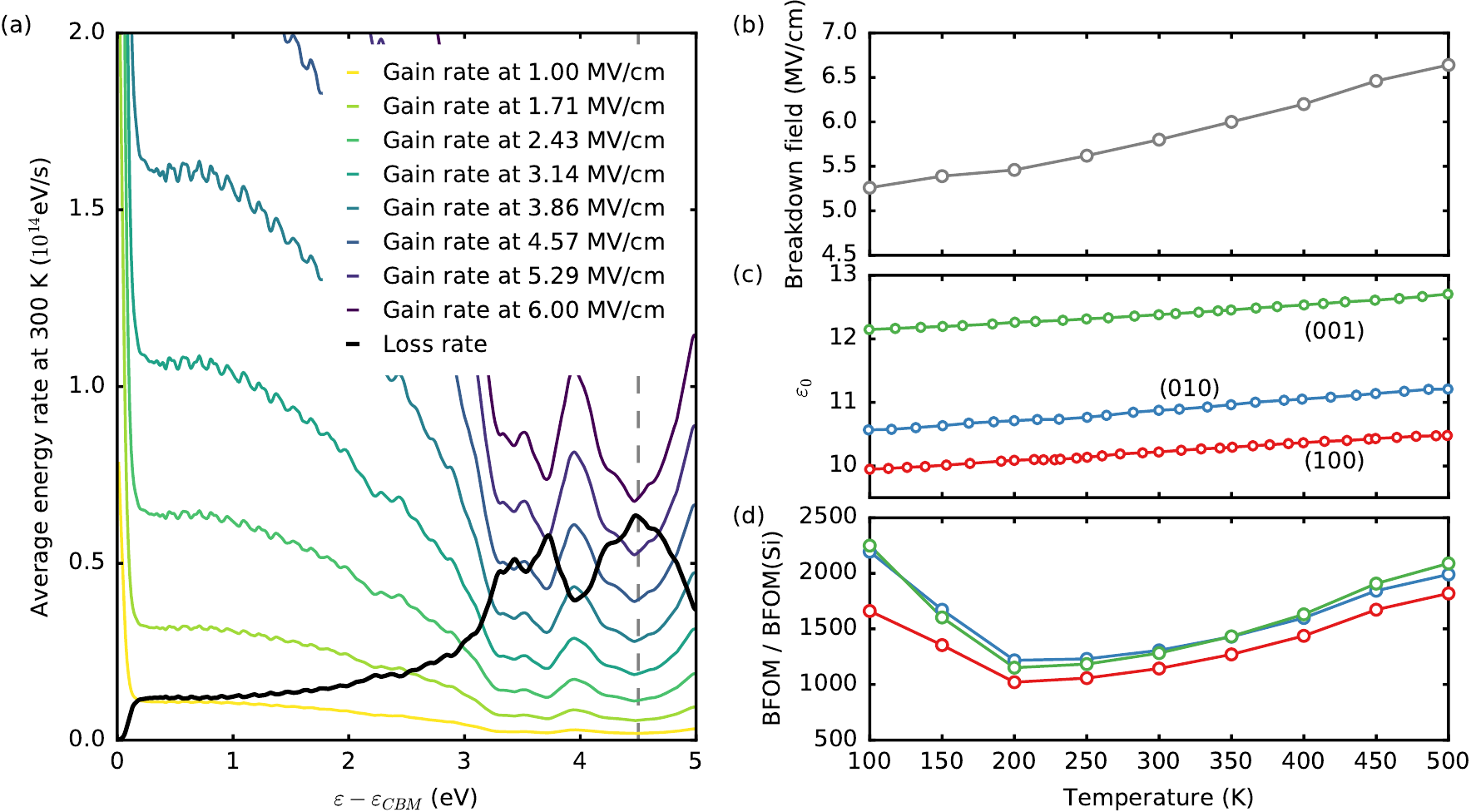}
  \caption{\label{fig6}
   (a) Average energy gain rate $A(\varepsilon)$ from an applied external field, and average energy loss to the lattice $B(\varepsilon)$. Both quantities are for 300~K.
   The intrinsic breakdown occurs when the applied electric field is such that the gain rate is larger than the loss rate for all energies between the conduction band minimum (CBM) and the CBM plus the energy of the bandgap (4.5 eV). 
   (b) Variation of computed breakdown field with temperature.  
   (c) Experimental variation of dielectric function with direction and temperature of $\beta$-Ga$_2$O$_3$ from Ref.~\cite{Fiedler2019}.
   (d) Baliga's figure of merit (BFOM) with respect to the BFOM of Silicon. The BFOM of Si was obtained using dielectric constants from Ref.~\cite{Karch1996} and a breakdown field of 0.3~MV/cm, as well as the experimental electron mobility from Norton~\cite{Norton1973}.   
  }
\end{figure*}

The BFOM is given by 
\begin{equation}\label{eq:bfom}
\textrm{BFOM} = \epsilon^{0} \mu E_b^3,
\end{equation}
 where $E_b$ is the computed breakdown field, $\mu$ the computed mobility from Eq.~\eqref{eq:btemob} and $\epsilon^{0}$ is the temperature-dependent experimental static dielectric function with the field perpendicular to the (100), (010) and (001) direction, respectively~\cite{Fiedler2019}, which we reproduce in Fig.~\ref{fig6}(c). 
Importantly, we stress that all the quantities entering in Eq.~\eqref{eq:bfom} are temperature-dependent.

The temperature and direction-dependent mobility has already been obtained in Section~\ref{sec:mob}. 
Therefore we only need to compute the breakdown field to obtain the BFOM. 
Refs.~\cite{Kim2016,Gorai2019} proposed the following model:
\begin{equation}
E_b = 24.442 \exp(0.315 \sqrt{E_g \omega_{\rm max}}),
\end{equation}   
where $E_g$ is the bandgap of the materials in eV, $\omega_{\rm max}$ the phonon cutoff frequency in THz and $E_b$ the breakdown field in MV/m.  
Although successful, the main limitation of this model is that it is independent of temperature. 
For this reason, we aim at computing the BFOM from first principles while retaining the temperature-dependence. 
To do so, in addition to the intrinsic carrier mobility, we need to compute the intrinsic breakdown field.

The most common theory for a material breakdown rely on electron avalanche~\cite{Sun2013} which occurs when the electron energy reaches the threshold for impact ionization. 
This is the energy at which an electron generates a second conduction electron by excitation across the electronic energy gap, causing electron multiplication (avalanche) and leading to a breakdown of the material~\cite{Sparks1981}. 
As a result, the threshold for impact ionization is usually taken as the electronic bandgap. 
The idea behind the theory relies on accelerating the conduction electron with a laser field and taking into account the electron scattering with the lattice during pumping. 
Indeed the phonon collision reduce the acceleration of the electron by modifying their momentum.

The von Hippel low energy criterion is more stringent and states that breakdown will occur when the rate of energy gain $A(E,\varepsilon,T)$ by an electron of energy $\varepsilon$ due to the external field $E$ at temperature $T$ is larger than the energy-loss rate $B(\varepsilon,T)$ to the lattice due to electron-phonon interaction~\cite{VonHippel1937,Sparks1981,Sun2012a}:
\begin{equation}
A(E,\varepsilon,T) > B(\varepsilon,T),
\end{equation} 
for energies $\varepsilon$ going from the conduction band minimum to the threshold for impact ionization, i.e. the bandgap of the materials.

The steady-state solution for the average energy-gain rate from the electric field is~\cite{Sparks1981}:
\begin{equation}
A(E,\varepsilon,T) = \frac{1}{3} \frac{e^2 \tau(\varepsilon,T)}{m^*} E^2
\end{equation}
where $e^2$ is the electron charge, $m^* = 0.3$~\cite{Ghosh2017} the electron effective mass. 
The energy and temperature-dependent electron-phonon lifetime is given by:
\begin{equation}
\tau^{-1}(\varepsilon,T) = \sum_{n\mathbf{k}} \tau_{n\mathbf{k}}^{-1}(T) \delta(\varepsilon_{n\mathbf{k}}-\varepsilon) / D(\varepsilon),
\end{equation} 
where $D(\varepsilon)$ is the density of state and $\tau_{n\mathbf{k}}^{-1}$ is given by Eq.~\eqref{eq:scattering_rate}.

The field-independent net rate of energy loss $B(\varepsilon, T)$ to the lattice is obtained by subtracting the rate of phonon absorption from phonon emission~\cite{Sparks1981,Sun2012a}:
\begin{align}\label{eq:enloss}
B(\varepsilon, T) =& \frac{2\pi}{\hbar D(\varepsilon)} \sum_{nm\nu} \!\iint\! \frac{d^3 k d^3 q}{\Omega_{\text{BZ}}^2} | g_{mn\nu}(\mathbf{k,q})|^2  \delta( \varepsilon_{n\mathbf{k}}-\varepsilon ) \nonumber \\
  \times & \hbar\omega_{\mathbf{q}\nu} \big[ (n_{\mathbf{q}\nu}  + 1/2)\delta(\varepsilon_{n\mathbf{k}} - \varepsilon_{m\mathbf{k+q}}  +  \hbar \omega_{\mathbf{q}\nu}) \nonumber \\
   - &  n_{\mathbf{q}\nu} \delta( \varepsilon_{n\mathbf{k}} - \varepsilon_{m\mathbf{k+q}}   -  \hbar \omega_{\mathbf{q}\nu})\big],
\end{align}
where $n_{\mathbf{q}\nu}$ are the Bose-Einstein occupation factors in the absence of an electric field.
%
%

We computed the energy gain and energy loss rates using the EPW software by interpolation on a dense 80$\times$80$\times$60 $\mathbf{k}$-point grid and a 40$\times$40$\times$30 $\mathbf{q}$-point grid with a constant smearing of 20~meV.
In Fig.~\ref{fig6}(a) we present the change of energy loss rate as a function of energy, starting from the conduction band minimum (CBM). 
On the same figure, we compare the loss rate with the average energy gain rate for increasing external electric field $E$ at room temperature. 
We define the intrinsic breakdown field $E_b$ as the smallest external electric field such that the energy gain curve is larger than the energy loss curve for all energies between the CBM and the CBM plus the energy of the bandgap (4.5~eV).
This value provides an estimate of the electric field range for which the material will not undergo dielectric breakdown.
We compute that at room temperature the breakdown field is 5.8~MV/cm including all electron-phonon scattering processes. 
Using the same approach for different temperatures, we can obtain the change of breakdown field with temperature shown in Fig.~\ref{fig6}(b). We find a breakdown field of 6.64~MV/cm at 500~K.

Such calculation was performed by Mengle and Kioupakis~\cite{Mengle2019} for the intrinsic electron breakdown field at 300~K. 
They obtained 5.4~MV/cm by considering only the dominant LO phonon mode, and estimated that the contribution of other modes would lead to 6.8~MV/cm.  
We note that the experimental breakdown field in $\beta$-Ga$_2$O$_3$ is typically reported around 8~MV/cm~\cite{Galazka2018}.
This is in line with our calculations, since the von Hippel low energy criterion should be seen as a lower bound for the breakdown field.

Using this information and the experimental dielectric function, we can compute the temperature and direction-dependent BFOM. 
The BFOM is typically given with respect to the BFOM of Silicon. 

In this case we computed the reference BFOM of Silicon by using the temperature-dependence dielectric constant of Refs.~\cite{Icenogle1976,Karch1996} and breakdown field of 0.3~MV/cm as well as the experimental temperature-dependent electron mobility from Norton~\cite{Norton1973}.
The resulting change of BFOM is given in Fig.~\ref{fig6}(d). The direction-averaged minimum and maximum values are 1130 and 2035, respectively.  
We see that even though the computed breakdown field underestimates the experiment,
this effect is compensated by an overestimation of the mobility.
As a result, our calculated BFOM is close to experimental estimates of 2000-3000~\cite{Galazka2018}.    

This cancellation suggests that the current level of theory could be sufficient to predict the BFOM of new materials.

\section{Conclusion}
In this work, we performed an in-depth study of the structural, vibrational, elastic, electrical, and transport properties of $\beta$-Ga$_2$O$_3$ using state-of-the art first-principles simulation tools. 
We carefully analyzed the structural properties of the monoclinic variation of $\beta$-Ga$_2$O$_3$ and analyzed the effect of spin-orbit coupling on those properties. 
We studied the electronic structure and carrier effective masses. 
We made a careful analysis of the vibrational properties including a symmetry analysis of $\beta$-Ga$_2$O$_3$ using first-order response function theory including dielectric and Born effective charges study.
We calculated many elastic properties by computing the elastic constants tensor including bulk, shear and Young modulus tensor using parametric three dimensional visualization  but also Poisson's ratio, universal elastic anisotropy, sound velocities and Debye temperature and found a strong directional anisotropy.         
We use the Boltzmann transport equation to compute the intrinsic electron and hole drift mobility and obtain a room temperature values of 258~cm$^2$/Vs and 1.2~cm$^2$/Vs, respectively. 
We found that the mobility in $\beta$-Ga$_2$O$_3$ was limited by a series of longitudinal optic phonons with symmetry character B$_u$ at the zone center. 
Finally we used the von Hippel low energy criterion to compute fully from first-principles the breakdown field which allowed us to compute the direction and temperature-dependent Baliga's figure of merit for high power device. 
We saw that the predicted figure of merit was in good agreement with experiment and attributed this to an overestimation of the computed mobility compensating an underestimation in the computed breakdown field.

The present analysis may serve as the basis for a general, consistent, and predictive framework to study materials for power electronics from first principles.

\section{Acknowledgement}
The authors thank Paolo Giannozzi for his help with the monoclinic structure in Quantum Espresso
and Andrea Dal Corso for his help with the monoclinic structure in the \texttt{thermo\_pw} code.    
Computer time was provided by the PRACE-15 and PRACE-17 resources MareNostrum at BSC-CNS, and the Texas Advanced Computing Center (TACC) at the University of Texas at Austin.
S.P. acknowledge support from the European Unions Horizon 2020 Research and Innovation Programme, under the Marie Sk\l{}odowska-Curie Grant Agreement SELPH2D No.~839217.
F.G.'s contribution to this work was supported as part of the Computational Materials Sciences Program funded by the U.S. Department of Energy, Office of Science, Basic Energy Sciences, under Award DE-SC0020129.

\section{Appendix}

\begin{table}[h!]
  \begin{tabular}{l r r }
  \toprule\\
Label    &  \multicolumn{2}{c}{Coordinates}   \\  
         & Peelaers~\cite{Peelaers2015}                 & This work   \\
\hline 
$N$      & (0,$\frac{1}{2}$,0)                          & (-$\frac{1}{2}$,0,0)                         \\
$X$      & ($1-\Psi,1-\Psi,0$)                          & ($\Psi-1$, $1-\Psi,0$)                       \\
$\Gamma$ & (0,0,0)                                      & (0,0,0)                                      \\
$M$      & (0,$\frac{1}{2}$,$\frac{1}{2}$)              & (-$\frac{1}{2}$,0,$\frac{1}{2}$)             \\
$I$      & ($\phi-1,\phi,\frac{1}{2}$)                  & (-$\phi$,$\phi-1,\frac{1}{2}$)               \\
$L$      & (-$\frac{1}{2}$,$\frac{1}{2}$,$\frac{1}{2}$) & (-$\frac{1}{2}$,-$\frac{1}{2}$,$\frac{1}{2}$)\\
$F$      & ($\zeta-1,1-\zeta,1-\eta$)                   & ($\zeta-1$, $\zeta-1,1-\eta$)                \\
$Y$      & (-$\frac{1}{2}$,$\frac{1}{2}$,0)             & (-$\frac{1}{2}$,-$\frac{1}{2}$,0)            \\
$\Gamma$ & (0,0,0)                                      & (0,0,0)                                      \\
$Z$      & (0,0,$\frac{1}{2}$)                          & (0,0,$\frac{1}{2}$)                          \\
\hline
$\Psi$   & 0.734  & 0.7336  \\
$\phi$   & 0.742  & 0.7418 \\
$\zeta$  & 0.397  & 0.3971  \\
$\eta$   & 0.590  & 0.5895 \\
  \botrule     
  \end{tabular}
  \caption{\label{table3}
 Reciprocal space coordinates of the high-symmetry point in the Brillouin zone of $\beta$-Ga$_2$O$_3$. 
 $\Psi=\frac{3}{4} - b^2/(4a^2 \sin^2 \beta)$, $\phi = \Psi - (\frac{3}{4} - \Psi)\frac{a}{c} \cos \beta$, $\zeta = (2+\frac{a}{c}\cos \beta)/(4\sin^2 \beta)$, $\eta = \frac{1}{2}-2\zeta \frac{c}{a}\cos \beta$.  
  }
\end{table}

\newpage

\bibliography{Bibliography} 

\end{document}